\documentclass[pdflatex,sn-mathphys-num]{sn-jnl}

\usepackage{amssymb}
\usepackage{amsmath}
\allowdisplaybreaks
\usepackage{cases}
\usepackage{pslatex}
\usepackage{amsfonts,color,morefloats}
\usepackage{amssymb,amsmath,latexsym,amsthm}
\usepackage{hyperref}
\usepackage{multirow}
\usepackage{makecell}
\usepackage{amsfonts}
\usepackage{tikz}
\usetikzlibrary{mindmap}
\usetikzlibrary{chains}
\usepackage{indentfirst}
\usepackage{amsmath}
\usepackage{amsthm}
\usepackage{array}
\usepackage{amssymb}
\usepackage{color}
\usepackage{graphicx}
\usepackage{geometry}
\usepackage{verbatim}
\usepackage{framed}
\usepackage{bm}
\usepackage{arydshln}
\usepackage[mathscr]{eucal}
\usepackage{nicematrix}
\usepackage[title]{appendix}

\newcommand{\lcm}{{\rm lcm}}

\newcommand{\Z}{\mathbb{{Z}}}

\newcommand{\tr}{{\mathrm{Tr}}}

\newcommand{\gf}{{\mathbb{F}}}

\newcommand{\C}{{\mathcal{C}}}

\newcommand{\bc}{{\mathbf{c}}}

\newtheorem{theorem}{Theorem}
\newtheorem{lemma}{Lemma}
\newtheorem{proposition}{Proposition}
\newtheorem{corollary}{Corollary}

\newtheorem{remark}{Remark}

\begin{document}

\title[The covering radii of a class of antiprimitive BCH codes]{The covering radii of a class of antiprimitive BCH codes}


\author[1]{\fnm{Ketong} \sur{Ren}}\email{rktswjtu@163.com}

\author*[2]{\fnm{Haode} \sur{Yan}}\email{yanhd@hit.edu.cn}

\author[1]{\fnm{Cuiling} \sur{Fan}}\email{fcl@swjtu.edu.cn}

\affil[1]{\orgdiv{School of Mathematics}, \orgname{Southwest Jiaotong University}, \orgaddress{\city{Chengdu}, \postcode{611756}, \country{China}}}

\affil*[2]{\orgdiv{School of Science}, \orgname{Harbin Institute of Technology}, \orgaddress{\city{Shenzhen}, \postcode{10587}, \state{518055}, \country{China}}}


\abstract{The covering radius is a fundamental parameter of linear error-correcting codes and plays a central role in maximum likelihood decoding. BCH codes form an important class of linear codes, owing to their strong error-correcting capability and efficient encoding and decoding algorithms. However, compared with primitive BCH codes, the covering radii of antiprimitive BCH codes remain much less studied. In this paper, we study the covering radius of a specific family of antiprimitive BCH codes, namely \(\C_{(q,q^m+1,3,0)}\), for all prime powers \(q\) and positive integers \(m\). Our approach is based on the analysis of several systems of equations over finite fields. As a byproduct, we identify a subfamily within this family that is simultaneously MDS and quasi-perfect.}




\maketitle

\section{Introduction}\label{sec:Intr}
	Let $q = p^s$ for some prime $p$ and a positive integer $s$. Let $\mathbb{F}_{q^m}$ denote the finite field of order $q^m$, where $m \in \mathbb{Z}^{+}$. For any two vectors $x, x' \in \mathbb{F}_q^n$, we denote by $d(x, x')$ the Hamming distance between them. A $k$-dimensional vector subspace $\mathcal{C} \subseteq \gf_q^n$ is called an $[n,k]$ linear code over $\gf_q$. The minimum distance $d$ of $\mathcal{C}$ is defined as $$d(\C) = \min\{d(c_1, c_2):c_1, c_2 \in \mathcal{C},c_1\neq c_2\}.$$ The covering radius of $\C$, denoted by $\rho(\C)$, is defined to be the smallest integer $\rho$ such that $\gf_{q}^n$ is the union of the spheres of radius $\rho$ centered at the codewords of $\C$. Equivalently, 
	$$\rho(\C)=\max\limits_{x\in\gf_q^n}\min\limits_{c\in\C}d(x,c).$$
    The covering radius is a fundamental geometric invariant of a code, quantifying how efficiently the ambient Hamming space is covered by balls centered at the codewords. In contrast to the minimum distance, which captures a pairwise local property, the covering radius reflects the global worst-case proximity of vectors to the code. It admits several key interpretations: in source coding, it is the maximum distortion when each vector is represented by its nearest codeword; in minimum-distance decoding, it is the largest possible distance from a received word to the code; for linear codes, it equals the maximum coset leader weight and the smallest integer such that every syndrome can be expressed as a combination of at most that many parity-check columns. Moreover, the covering radius is intimately tied to code maximality, as maximal codes of prescribed length and minimum distance are characterized by bounds involving this parameter. Owing to these diverse connections, the covering radius is a central quantity in coding theory, with applications to decoding, data compression, testing, write-once memories, and related combinatorial covering problems~\cite{CRclctn,CR85to94,CRSurvey1985,Guru2003}.
    
    It is well known that $\rho(\C) \geqslant \lfloor \frac{d(\C)-1}{2} \rfloor$. A classic result links it to perfect codes: a code $\C$ is perfect if and only if $\rho(\C) = \lfloor \frac{d(\C)-1}{2} \rfloor$. Furthermore, a code is called quasi-perfect if $\rho(\C) = \lfloor \frac{d(\C)-1}{2} \rfloor+1$. Beyond these two well-structured classes of codes, determining the covering radius of a general linear code is a challenging problem. It has been proven in \cite{MAM84} that computing the exact covering radius of a code is both NP-hard and co-NP-hard. The covering radii of most linear codes are not determined exactly; prevalent results in the literature establish bounds on them (see, e.g., \cite{CRBound1}, \cite{CRBound8}, \cite{CRBound6}, \cite{CRBound9}, \cite{CRBound7}, \cite{CRBound2}, \cite{CRBound4}, \cite{CRBound3}, \cite{CRBound5}). Only a limited number of classes have known exact covering radii \cite{CRValue1}, \cite{CRValue6}, \cite{CRValue7}, \cite{CRValue3}, \cite{CR_BCH_Va_2}, \cite{CRValue9}, \cite{CR_MDS_1}, \cite{CRValue2}, \cite{CR_MelasCode_1}, \cite{CR_ZeBeGCode_1}, \cite{CR_ZeBeGCode_2}, \cite{CR_ZeBeGCode_3}, \cite{CRValue8}. Therefore, determining exact covering radii for explicit algebraically structured code families is of fundamental importance: it elucidates their metric geometry and reveals whether a code is perfect, quasi-perfect, or far from these extremal configurations. This problem is particularly significant for cyclic and BCH codes, whose algebraic structure reduces the covering-radius question to equations over finite fields, yielding exact values that capture information inaccessible from the minimum distance alone.
    
    The covering radius of a linear code can be characterized through its parity-check matrix. A parity check matrix $H$ of an $[n,k]$ code C is an $(n-k)\times n$ matrix with the property that a vector $\bc\in\gf_q^n$ is a codeword if and only if it satisfies the system of linear constraints: $H\cdot\bc^{\top}= 0$. The following lemma which will be used repeatedly states the relation between a parity-check matrix of a code and its covering radius.
    \begin{lemma}\label{lem:CR_PCmatrix}\cite[Theorem 11.1.2]{HuffmanPless03}
        If $\C$ is an $[n,k]$ linear code over $\gf_q$ with parity check matrix $H$, then $\rho(\C)$ is the smallest number such that every non-zero vector in $\gf_q^{n-k}$ is a combination of $\rho(\C)$ or fewer columns of $H$.
    \end{lemma}

    BCH codes form one of the most important families of linear codes, owing to their strong error-correcting capability and efficient encoding and decoding algorithms. Research on the covering radii of BCH codes has been limited and focused mainly on primitive BCH codes; see \cite{C97}, \cite{CR_BCH_Va_3}, \cite{DWN60}, \cite{CRValue4}, \cite{CR_BCH_2nd09}, \cite{CR_BCH_Bd_1}, \cite{CR_BCH_Va_1}, \cite{CR_BCH_2nd25} for more details. In contrast, antiprimitive BCH codes have received comparatively less attention from the viewpoint of covering radii.

    In this paper, we study the covering radii of the antiprimitive BCH codes $\C_{(q,q^m+1,3,0)},$ where \(q\) is a prime power and \(m\) is a positive integer. The paper is organized as follows. Section \ref{sec:Pre} introduces the notation and preliminary results on cyclic codes, BCH codes, quadratic equations over finite fields, and character sums. In Section \ref{sec:Sums}, we explicitly evaluate several character sums that are essential in the later arguments. With these foundations, in Sections \ref{sec:CR_meq1} and \ref{sec:CR_0} we compute the covering radii of the antiprimitive BCH codes $\mathcal{C}_{(q,q^m+1,3,0)}$. More precisely, Section \ref{sec:CR_meq1} treats the special case $m=1$, determining the covering radius of $\mathcal{C}_{(q,q+1,3,0)}$, while Section \ref{sec:CR_0} addresses the general case $m>1$. Finally, Section \ref{sec:Con} concludes the paper.

\section{Preliminaries}\label{sec:Pre}
    In this section, we collect the notation and preliminary results used throughout the paper. We first recall some standard facts on cyclic codes and BCH codes, then state the parity-check characterization of covering radius. We also recall several results on quadratic equations over finite fields and character sums.

\subsection{Cyclic codes and BCH codes}
    In the following, we always assume \(\gcd(n,q)=1\). By identifying any vector \((c_0,c_1,\dots,c_{n-1})\in \mathbb{F}_{q}^{n}\) with
        \[c_{0}+c_{1}x+c_{2}x^{2}+\cdots+c_{n-1}x^{n-1}\in \mathbb{F}_{q}[x]/(x^{n}-1),\]
    any code \(\mathcal{C}\) of length \(n\) over \(\mathbb{F}_{q}\) corresponds to a subset of the ring \(\mathbb{F}_{q}[x]/(x^{n}-1)\). A \(q\)-ary linear code \(\mathcal{C}\) of length \(n\) is called cyclic if \((c_{0},c_{1},\dots,c_{n-1})\in \mathcal{C}\) implies \((c_{n-1},c_{0},c_{1},\dots,c_{n-2})\in \mathcal{C}\). Then, a linear code \(\mathcal{C}\) is cyclic if and only if the corresponding subset in \(\mathbb{F}_{q}[x]/(x^{n}-1)\) is an ideal of \(\mathbb{F}_{q}[x]/(x^{n}-1)\). Note that each ideal of \(\mathbb{F}_{q}[x]/(x^{n}-1)\) is principal. Let \(\mathcal{C}=\langle g(x)\rangle\) be a cyclic code, where \(g(x)\) is monic and has the smallest degree among the generators of \(\mathcal{C}\). Then \(g(x)\) is called the generator polynomial of \(\mathcal{C}\), where \(g(x)\) is unique and a divisor of \(x^{n}-1\). The polynomial \(h(x)=(x^{n}-1)/g(x)\) is called the parity-check polynomial of \(\mathcal{C}\).

    Let \(\mathbb{Z}_{n}\) denote the set \(\{0,1,2,\dots,n-1\}\). Let \(b\) be an integer satisfying \(0\leqslant b<n\). The \(q\)-cyclotomic coset of \(b\) modulo \(n\) is defined as
        \[C_{b}=\{b,bq,bq^{2},\dots,bq^{l_{b}-1}\} \subseteq \mathbb{Z}_{n},\]
    where \(l_{b}\) is the smallest positive integer such that \(b\equiv bq^{l_{b}}\pmod n\). The integer \(l_{b}\) is called the size of the \(q\)-cyclotomic coset \(C_{b}\). The smallest integer in \(C_{b}\) is referred to as the coset leader of \(C_{b}\). Let \(\Gamma(n,q)\) denote the set of all coset leaders. For two distinct elements \(b,t\in\Gamma(n,q)\), we have \(C_{b}\cap C_{t}=\emptyset\), and furthermore,
        \[\bigcup_{b\in\Gamma(n,q)} C_{b}=\mathbb{Z}_{n}.\]
    Let \(e=\operatorname{ord}_{n}(q)\) denote the order of \(q\) modulo \(n\), which is the smallest positive integer such that \(q^{e}\equiv 1\pmod n\). Let \(\alpha\) be a primitive element in \(\mathbb{F}_{q^{e}}\) and define \(\beta=\alpha^{\frac{q^{e}-1}{n}}\). Then, \(\beta\) is a primitive \(n\)-th root of unity in \(\mathbb{F}_{q^{e}}\). The minimal polynomial \(M_{\beta^{b}}(x)\) of \(\beta^{b}\) over \(\mathbb{F}_{q}\) is the monic polynomial of the smallest degree over \(\mathbb{F}_{q}\) having \(\beta^{b}\) as a root. It can be expressed as
        \[M_{\beta^{b}}(x)=\prod_{i\in C_{b}}(x-\beta^{i})\in\mathbb{F}_{q}[x],\]
    which is irreducible over \(\mathbb{F}_{q}\). Consequently, we have
        \[x^{n}-1=\prod_{b\in\Gamma(n,q)} M_{\beta^{b}}(x).\]

    We are now ready to introduce BCH codes, which are a subclass of cyclic codes with many interesting properties. Let \(\delta\) be an integer satisfying \(2\leqslant \delta\leqslant n\) and \(h\) be an integer with $0\leqslant h <n$. A BCH code over \(\mathbb{F}_{q}\) with length \(n\) and designed distance \(\delta\), denoted by \(\mathcal{C}_{(q,n,\delta,h)}\), is a cyclic code with the generator polynomial
        \[g_{(q,n,\delta,h)}(x)=\operatorname{lcm}\bigl(M_{\beta^{h}}(x), M_{\beta^{h+1}}(x),\dots, M_{\beta^{h+\delta-2}}(x)\bigr),\]
    where \(\lcm(\cdot)\) denotes the least common multiple in the ring \(\mathbb{F}_{q}[x]\). If \(h=1\), the BCH code \(\mathcal{C}_(q,n,\delta,1)\) is called a narrow-sense BCH code. If \(n=q^{m}-1\) for some positive integer \(m\), then the code \(\mathcal{C}_(q,n,\delta,h)\) is called a primitive BCH code. If \(n=q^{m}+1\), it is called an antiprimitive BCH code by Ding in \cite{Ding2016}. For the specific antiprimitive length \(n=q^m+1\) considered in this paper, let \(\alpha\) be a primitive element of \(\mathbb{F}_{q^{2m}}\), and set $\mu=\alpha^{q^m-1}.$ Then \(\mu\) is a primitive \((q^m+1)\)-th root of unity. Define
        $$U_{q^m+1}=\{u\in\mathbb{F}_{q^{2m}} : u^{q^m+1} = 1\}.$$
    Since \(\mu\) has order \(q^m+1\), we have $U_{q^m+1}=\langle\mu\rangle$. With this notation, an extension-field parity-check matrix of \(\C_{(q,q^m+1,\delta,h)}\) is given by
        $$H_{(q,q^m+1,\delta,h)}=\begin{pmatrix}
                1 & \mu^{h} & \mu^{2h} & \cdots & \mu^{q^mh}  \\
                1 & \mu^{h+1} & \mu^{2(h+1)} & \cdots & \mu^{q^m(h+1)} \\
                \vdots & \vdots & \vdots & & \vdots \\
                1 & \mu^{h+\delta-2} & \mu^{2(h+\delta-2)} & \cdots & \mu^{q^m(h+\delta-2)} \\
            \end{pmatrix}.$$

\subsection{Notation}
For clarity, the key notations adopted throughout the paper are summarized in this subsection.
\begin{itemize}
	\item Let $s$ and $m$ be two positive integers.
	\item Let $p$ be a prime and $q=p^s$. Let $\mathbb{F}_{q^m}$ denote the finite field of order $q^m$.
	\item Let $\alpha$ be a primitive element of $\gf_{q^{2m}}$. Let $\mu=\alpha^{q^m-1}$, then $\mu$ is the primitive $(q^m+1)$-th root of unity.
	\item For arbitrary $x\in\gf_{q^{2m}}$, denote by $\overline{x}$ the $q^m$-th power of $x$, i.e., $\overline{x}=x^{q^m}$. 
	\item Define $U_{q^m+1}=\{u\in\gf_{q^{2m}}:u\overline{u}=1\}=\langle\mu\rangle$.
	\item Denote the covering radius of a linear code $\C$  by $\rho(\C)$.
	\item For the $q$-ary BCH code $\C_{(q,q^m+1,3,0)}$ over \(\gf_q\), denote by $H_{(q,q^m+1,3,0)}$ the following \(2\times(q^m+1)\) compact parity-check matrix over \(\gf_{q^{2m}}\):
            $$H_{(q,q^m+1,3,0)}=\begin{pmatrix}
                1 & 1 & 1 & \cdots & 1 \\
                1 & \mu^{1} & \mu^{2} & \cdots & \mu^{q^m} 
            \end{pmatrix}.$$
        Equivalently, by keeping the first row as a row over \(\gf_q\) and expanding the entries of the second row with respect to a basis of \(\gf_{q^{2m}}\) over \(\gf_q\), this matrix induces a \((2m+1)\times(q^m+1)\) parity-check matrix over \(\gf_q\).
\end{itemize}

\subsection{A parametrization of \(U_{q^m+1}\)}
The next lemma gives a rational parametrization of \(U_{q^m+1}\setminus\{1\}\). It will be used in the sequel to convert character sums over \(U_{q^m+1}\) into character sums over \(\mathbb F_{q^m}\).
\begin{lemma}\label{lem:U_to_F}\cite{Tzr2018}
    Let $ A \in \gf_{q^{2m}}\setminus\gf_{q^m}$ be fixed. Then
    $$U_{q^m+1}\setminus\{1\}=\left\{\frac{u+\overline{A}}{u+A} : u\in\gf_{q^m}\right\}.$$
\end{lemma}

\subsection{Quadratic equations over finite fields}
This subsection summarizes some results on quadratic equations over finite fields. We begin by recalling the definition of the trace function. Let \(m'\) be a positive divisor of \(m\). The trace function from \(\mathbb{F}_{q^m}\) to \(\mathbb{F}_{q^{m'}}\) is defined as
    $$\tr_{m'}^{m}=x+x^{q^{m'}}+\dots+x^{q^{\left(\frac{m}{m'}-1\right)m'}}.$$
The following lemma provides conditions under which a quadratic equation over a finite field of even characteristic has solutions in $U_{q^m+1}$.
\begin{lemma}\label{lem:sol_in_unit_peq2}\cite{Tzr2018}
	Let $n=2m$ be an even positive integer and $a,b\in \gf^*_{2^n}$ satisfying $\mathrm{Tr}_1^n\left(\frac{b}{a^2}\right)=0$.  Then the quadratic equation $x^2+ax+b=0$ has
	\begin{itemize}
		\item both solutions in $U_{q^m+1}$ if and only if
			$$b=\frac{a}{\overline{a}},~\mathrm{Tr}_1^m\left(\frac{b}{a^2}\right)=\mathrm{Tr}_1^m\left(\frac{1}{a\overline{a}}\right)=1.$$
		\item exactly one solution in $U_{q^m+1}$ if and only if
			$$b\neq \frac{a}{\overline{a}},~(1+b\overline{b})(1+a\overline{a}+b\overline{b})+a^2\overline{b}+\overline{a}^2b=0.$$
	\end{itemize}
\end{lemma}

The following lemma provides conditions under which a quadratic equation over a finite field of odd characteristic has solutions in $U_{q^m+1}$.
\begin{lemma}\label{lem:sol_in_unit_p}\cite{Tzr2019}
	Let $p\geqslant3$ be an odd prime, $n=2m$ for a positive integer $m$ and $a,b\in \gf^*_{p^n}$ such that $\Delta=a^2-4b$ is a square in $\gf_{p^n}$. Then for the quadratic equation $x^2+ax+b=0$, we have
	\begin{itemize}
		\item both solutions in $U_{q^m+1}$, if and only if 
				$$b=\frac{a}{\overline{a}},~\theta+\overline{\theta}=0 \mbox{ where } \theta=\sqrt{\frac{a\overline{a}-4}{a\overline{a}}}.$$
		\item there is exactly one solution in $U_{q^m+1}$, if and only if 
				$$b\neq\frac{a}{\overline{a}},~2+2b\overline{b}-a\overline{a}-\sqrt{\Delta}\overline{\sqrt{\Delta}}=0.$$
	\end{itemize}
\end{lemma}

\subsection{Character sums over finite fields}
In this subsection, we recall some basic facts on character sums over finite fields. The additive character and multiplicative character used in this paper are defined below.

\subsubsection*{Additive character}
An additive character over \(\gf_{p^m}\) is a function $\psi:\mathbb{F}_{p^m} \rightarrow \mathbb{C}^\times$ such that for all $\beta_1,\beta_2\in\mathbb{F}_{p^m}$,
    $$\psi(\beta_1+\beta_2)=\psi(\beta_1)\,\psi(\beta_2).$$
The trace function \(\tr_1^m\) from $\gf_{p^m}$ to $\gf_p$ provides a canonical construction of an additive character. The trace function is linear; that is, for all \(\beta_1, \beta_2 \in \mathbb{F}_{p^m}\) and any $c_1,c_2 \in \mathbb{F}_p$,
    $$\tr_1^m(c_1\beta_1 + c_2\beta_2) = c_1 \tr_1^m(\beta_1) + c_2 \tr_1^m(\beta_2).$$
Let \(\zeta_p = e^{\frac{2\pi i}{p}}\) be a primitive \(p\)-th root of unity. Then the function
    $$\psi(\beta) = \zeta_p^{\tr_1^m(\beta)}$$
is an additive character of \(\mathbb{F}_{p^m}\).

As a powerful tool, bounds on character sums (e.g., the Carlitz-Uchiyama bound \cite{FF}) provide key estimates. In particular, a variant where the summation runs over characters of rational functions \cite[Lemma 1]{CR_BCH_2nd25} will be utilized in this work.

\begin{lemma}\cite{TC06}\label{lem:WeilBound_add_rational}
	Consider the finite field $\gf_q$ with characteristic $p$. Let $f(X)\in \gf_q(X)$ be a rational function, 
	\[f(X)=P(X)+\frac{R(X)}{Q(X)}, \quad \deg(R(X))<\deg(Q(X)), \quad M:=\deg(P(X)),\]
	where $P(X),Q(X),R(X)\in \gf_q[x]$ are polynomials, and $\gcd(R(X),Q(X))=1$. Suppose further that $f(X)$ is non-constant, and that $M=0$ or $p\nmid M$. Write
	\[Q(X)=\prod_{i=1}^tQ_i(X)^{m_i},\]
	where $Q_i(X)$ are distinct irreducible polynomials over $\gf_q$, and $m_i\geq 1$ for each $i$. Define 
	\[L:=\sum_{i=1}^t(m_i+1)\deg(Q_i(X)).\]
	Then there exist complex numbers $\omega_j\in\mathbb{C}$, $1\leq j \leq M+L-1$, such that
	\[\sum_{\beta\in\gf_q\setminus S}\psi(f(\beta))=-\sum_{j=1}^{M+L-1}\omega_j,\]
	where $S$ is the set of poles of $f(X)$. Additionally, $\left|\omega_j\right|=\sqrt{q}$ for all $j$, except for a single value, $j'$, satisfying $|\omega_{j'}|=1$. Thus, 
	\[\left|\sum_{\beta\in\gf_q\setminus S}\psi(f(\beta))\right| \leq 1+\left(M+L-2\right)\sqrt{q}.\]
\end{lemma}

The following estimate is a useful consequence of Lemma \ref{lem:WeilBound_add_rational} and the parametrization of \(U_{q^m+1}\) given above. It will be used later in the treatment of even-characteristic cases.
\begin{lemma}\label{lem:addsum_uB_on_u}
    Let $p=2$ and $q=2^s$. Let $E_1\in\gf_{q^m}$ and $E_0, E_2 \in\gf_{q^{2m}}^*$ satisfy $E_0=\overline{E_2}$ and $E_0+E_1+E_2\neq0$. Denote by $R$ the set of zeros of $E_2x^2+E_1x+E_0=0$ in $U_{q^m+1}$. (Note that $1 \notin R$ by the assumption $E_0+E_1+E_2 \neq 0$.) Then we have the estimate
        $$\sum\limits_{x\in U_{q^m+1}\setminus R}(-1)^{\tr^{sm}_1\left(\frac{x}{E_2x^2+E_1x+E_0}\right)}\leqslant 2+2\sqrt{q^m}.$$
\end{lemma}
\begin{proof}
    By Lemma \ref{lem:U_to_F}, for any given $A$ in $\gf_{q^{2m}}\setminus\gf_{q^m}$, we have
    $$\sum\limits_{x\in U_{q^m+1}\setminus R}(-1)^{\tr^{sm}_1\left(\frac{x}{E_2x^2+E_1x+E_0}\right)}=(-1)^{\tr^{sm}_1\left(\frac{1}{E_0+E_2+E_1}\right)}+\sum\limits_{\gamma\in \gf_{q^m}\setminus R_{\gamma}}(-1)^{\tr^{sm}_1\left(\frac{(\gamma+\overline{A})(\gamma+A)}{(E_0+E_1+E_2)\gamma^2+E_1(A+\overline{A})\gamma+E_2\overline{A}^2+E_1\overline{A}A+E_0A^2}\right)},$$
    where $R_{\gamma}=\{\gamma\in\gf_{q^m}:(E_0+E_1+E_2)\gamma^2+E_1(A+\overline{A})\gamma+E_2\overline{A}^2+E_1\overline{A}A+E_0A^2=0\}.$
    By Lemma \ref{lem:WeilBound_add_rational}, we have $M = 0$ and $L \leqslant 4$, which gives
    $$\sum\limits_{\gamma\in \gf_{q^m}\setminus S_{\gamma}}(-1)^{\tr^{sm}_1\left(\frac{(\gamma+\overline{A})(\gamma+A)}{(E_0+E_1+E_2)\gamma^2+E_1(A+\overline{A})\gamma+E_2\overline{A}^2+E_1\overline{A}A+E_0A^2}\right)}\leqslant 1+2\sqrt{q^m}.$$
    The proof is complete.
\end{proof}

\subsubsection*{Multiplicative character}
A multiplicative character over \(\gf_{p^m}\) is a function $\chi:\gf_{p^m}^* \rightarrow \mathbb{C}^*$ such that for all \(\beta_1,\beta_2\in \gf_{p^m}^*\),
    $$\chi(\beta_1\beta_2)=\chi(\beta_1)\chi(\beta_2).$$
Equivalently, a multiplicative character is a group homomorphism from the cyclic group \(\gf_{p^m}^*\) to \(\mathbb{C}^{*}\). If \(\chi\) is not the trivial character, we extend it to the whole field \(\gf_{p^m}\) by setting \(\chi(0)=0\). The order of a multiplicative character \(\chi\) is the smallest positive integer \(r\) such that \(\chi^r\) is the trivial character.

In particular, when \(p\) is odd, the unique multiplicative character of order \(2\) over \(\gf_{p^m}\) is called the quadratic character. We denote it by \(\chi_{p^m}(\cdot)\), and it is given by
$$\chi_{p^m}(x)=
\left\{
\begin{array}{cl}
	1, & \mbox{if } x \mbox{ is a square in } \gf_{p^m}^*,\\
	-1,& \mbox{if } x \mbox{ is a nonsquare in }\gf_{p^m}^*, \\
	0, & \mbox{if } x=0.
\end{array}
\right.
$$
Let $\gf_{p^m}[x]$ be the polynomial ring over $\gf_{p^m}$. We consider the character sum of the form
\begin{eqnarray} \label{2:x1} \sum_{c\in\gf_{p^m}}\chi_{p^m}(f(c))\end{eqnarray}
with $f\in\gf_{p^m}[x]$. The case of $\deg(f) =1$ is trivial, and for $\deg(f) = 2$, the exact value is given by the following lemma.

\begin{lemma}\cite[Theorem 5.48]{FF}\label{lem:chsum_deg2}
	Let $f(x)=a_2x^2+a_1x+a_0\in\gf_q[x]$ with $q$ odd and $a_2\neq 0$. Put $d=a_1^2-4a_0a_2$ and let $\chi(\cdot)$ be the quadratic character of $\gf_q$. Then
	$$\sum\limits_{c\in\gf_q}\chi_q(f(c))=
	\left\{
	\begin{array}{lcl}
		-\chi(a_2), & \mbox{if }d\neq 0,\\
		(q-1)\chi(a_2), & \mbox{if }d=0. \\
	\end{array}
	\right.
	$$
\end{lemma}
For $\deg(f) > 2$, we have
\begin{lemma}\cite[Theorem 5.41]{FF}\label{lem:WeilBound_multi}
    Let $\chi$ be a multiplicative character of $\gf_q$ of order $r>1$ and let $f\in\gf_q[x]$ be a monic polynomial of positive degree that is not an $m$-th power of a polynomial. Let $d$ be the number of distinct roots of $f$ in its splitting field over $\gf_q$. Then for every $a\in\gf_q$, we have
        $$\left|\sum\limits_{c\in\gf_q}\chi(af(c))\right|\leqslant(d-1)q^{\frac{1}{2}}.$$
\end{lemma}

\section{auxiliary character sums and counting results}\label{sec:Sums}
This section is devoted to several auxiliary computations that will be used in Sections \ref{sec:CR_meq1} and \ref{sec:CR_0}. In addition to estimating certain character sums over finite fields and over the unit circle \(U_{q^m+1}\), we also count tuples of elements in \(U_{q^m+1}\) satisfying prescribed conditions. These results provide the technical ingredients needed to analyze the systems of equations obtained from the parity-check characterization of the covering radius. 

We first record several identities and estimates in even characteristic. Lemma \ref{lem:values_for_addsum} gives a trace identity on \(U_{q^m+1}\), Lemma \ref{lem:meq1_qeven_chsum} evaluates two sums needed for the case \(m=1\), and Lemma \ref{lem:addsum_for_CR} provides counting formulas and character-sum estimates used in the case \(m>1\).

\begin{lemma} \label{lem:values_for_addsum}
    Let $p=2$. If $u\in U_{q^m+1}\setminus\{1\}$, then it holds that $\tr^{sm}_1\left(\frac{1}{u+u^{-1}}\right)=1.$
\end{lemma}
\begin{proof}
   Since $u$ and $u^{-1}$ are two distinct solutions in $U_{q^m+1}$ of $x^2+(u+u^{-1})x+1=0$, the necessary and sufficient condition in Lemma \ref{lem:sol_in_unit_peq2} immediately yields $\tr_1^{sm}\bigl(1/(u+u^{-1})^2\bigr)=\tr_1^{sm}\bigl(1/(u+u^{-1})\bigr)=1$.
\end{proof}

The following lemma is used to determine $\rho\bigl(\mathcal{C}_{(q,q+1,3,0)}\bigr)$.
\begin{lemma}\label{lem:meq1_qeven_chsum}
    Let $q$ be a power of $2$ and let $\beta\in\gf_{q^2}\setminus\{1\}$. Write \(\overline{\beta}=\beta^q\). Then the following holds:
    \begin{itemize}
        \item $$\lvert \{(a_1,a_2)\in(\mathbb F_q^*)^2:a_1+a_2+\beta+1\ne0,\beta\overline{\beta}+(\beta+\overline{\beta})(a_1+a_2+1)+1\ne0\}\rvert =q^2-3q+3.$$
        \item $$\sum\limits_{\substack{a\in\gf_q^*\\\beta\overline{\beta}+(\beta+\overline{\beta})(a+1)+1\neq0}}(-1)^{\tr^s_1\left(\frac{a^2(a^2+(\beta+\overline{\beta})a+(\beta+1)(\overline{\beta}+1))}{(\beta\overline{\beta}+(\beta+\overline{\beta})(a+1)+1)^2}\right)}=\begin{cases}
            q-1, &\text{ when } \beta\in\gf_q\setminus\{1\},\\
            q-2, &\text{ when } \beta\not\in\gf_q.
        \end{cases}$$
        \item $$\sum\limits_{\substack{a_1,a_2\in\gf_q\\\beta\overline{\beta}+(\beta+\overline{\beta})(a_1+a_2+1)+1\neq0}}(-1)^{\tr^s_1\left(\frac{a_2^2(a_1+a_2+\beta+1)(a_1+a_2+\overline{\beta}+1)}{(\beta\overline{\beta}+(\beta+\overline{\beta})(a_1+a_2+1)+1)^2}\right)}=\begin{cases}
            q, &\text{ when } \beta\in\gf_q\setminus\{1\},\\
            0, &\text{ when } \beta\not\in\gf_q,
        \end{cases}$$
    \end{itemize}
\end{lemma}
\begin{proof} We prove the equalities in turn, as detailed below.
    \begin{itemize}
        \item First assume that \(\beta\in\mathbb F_q\). Then \(\overline{\beta}=\beta\), and since \(q\) is even, we have $\beta+\overline{\beta}=0.$ Hence the second inequality becomes
            $$\beta\overline{\beta}+1=\beta^2+1=(\beta+1)^2\ne0,$$
        because \(\beta\ne1\). Therefore, in this case only the first inequality excludes pairs. Therefore,
        \begin{align*}
            &\lvert \{(a_1,a_2)\in(\mathbb F_q^*)^2:a_1+a_2+\beta+1\ne0,\beta\overline{\beta}+(\beta+\overline{\beta})(a_1+a_2+1)+1\ne0\}\rvert\\
            =&\sum\limits_{a_1\in\gf_q^*}\sum\limits_{a_2\in\gf_q^*\setminus\{a_1+\beta+1\}}1\\
            =&(q-1)+\sum\limits_{a_1\in\gf_q^*\setminus\{\beta+1\}}\sum\limits_{a_1\in\gf_q^*\setminus\{a_1+\beta+1\}}1\\
            =&q^2-3q+3.
        \end{align*}
        Next assume that \(\beta\notin\mathbb F_q\). Then the first inequality excludes no pair, since \(a_1+a_2+1\in\mathbb F_q\) for all \(a_1,a_2\in\mathbb F_q\), whereas \(\beta\notin\mathbb F_q\). Moreover, \(\beta+\overline{\beta}\ne0\). Thus the second inequality excludes exactly the pairs satisfying $\beta\overline{\beta}+(\beta+\overline{\beta})(a_1+a_2+1)+1=0,$ or equivalently,
            $$a_1+a_2=\frac{\beta\overline{\beta}+1}{\beta+\overline{\beta}}+1.$$
        The element \(\frac{\beta\overline{\beta}+1}{\beta+\overline{\beta}}+1\) belongs to \(\mathbb F_q\) and is nonzero, because \(\frac{\beta\overline{\beta}+1}{\beta+\overline{\beta}}+1=0\) would give \((\beta+1)(\overline{\beta}+1)=0\), contradicting \(\beta\ne1\). Therefore,
        \begin{align*}
            &\lvert \{(a_1,a_2)\in(\mathbb F_q^*)^2 : a_1+a_2+\beta+1\ne0,\; \beta\overline{\beta}+(\beta+\overline{\beta})(a_1+a_2+1)+1\ne0\}\rvert \\
            =&\lvert \{(a_1,a_2)\in(\mathbb F_q^*)^2 : a_1+a_2\ne \frac{\beta\overline{\beta}+1}{\beta+\overline{\beta}}+1\}\rvert \\
            =&\sum_{a_1\in\mathbb F_q^*}\sum_{a_2\in\mathbb F_q^*\setminus\{a_1+\frac{\beta\overline{\beta}+1}{\beta+\overline{\beta}}+1\}} 1 \\
            =&(q-1)+\sum_{a_1\in\mathbb F_q^*\setminus\{\frac{\beta\overline{\beta}+1}{\beta+\overline{\beta}}+1\}}\sum_{a_2\in\mathbb F_q^*\setminus\{a_1+\frac{\beta\overline{\beta}+1}{\beta+\overline{\beta}}+1\}} 1 \\
            =&q^2-3q+3.
        \end{align*}
        \item For any $\beta\in\gf_{q^2}\setminus\{1\}$, we have
            \begin{align*}
                &\sum\limits_{\substack{a\in\gf_q^*\\\beta\overline{\beta}+(\beta+\overline{\beta})(a+1)+1\neq0}}(-1)^{\tr^s_1\left(\frac{a^2(a^2+(\beta+\overline{\beta})a+(\beta+1)(\overline{\beta}+1))}{(\beta\overline{\beta}+(\beta+\overline{\beta})(a+1)+1)^2}\right)}\\
                =&\sum\limits_{\substack{a\in\gf_q^*\\(\beta+\overline{\beta})a+(\beta+1)(\overline{\beta}+1)\neq0}}(-1)^{\tr_1^s\left(\frac{a^4}{((\beta+\overline{\beta})a+(\beta+1)(\overline{\beta}+1))^2}\right) + \tr^s_1\left(\frac{a^2((\beta+\overline{\beta})a+(\beta+1)(\overline{\beta}+1))}{((\beta+\overline{\beta})a+(\beta+1)(\overline{\beta}+1))^2}\right)} \\
                =&\sum\limits_{\substack{a\in\gf_q^*\\(\beta+\overline{\beta})a+(\beta+1)(\overline{\beta}+1)\neq0}}(-1)^{\tr^s_1\left(\frac{a^2}{(\beta+\overline{\beta})a+(\beta+1)(\overline{\beta}+1)}\right)+\tr^s_1\left(\frac{a^2}{(\beta+\overline{\beta})a+(\beta+1)(\overline{\beta}+1)}\right)}\\
                =&\left\lvert \{a : a \in \gf_q^* \text{ and } (\beta+\overline{\beta})a+(\beta+1)(\overline{\beta}+1)\neq0\}\right\rvert.
            \end{align*}
            We now evaluate this sum by distinguishing the possible values of $\beta$.
            \begin{itemize}
                \item When $\beta\not\in\gf_q$, $\beta+\overline{\beta}\neq0$ and $\beta\neq1$, $(\beta+\overline{\beta})a+(\beta+1)(\overline{\beta}+1)\neq0$ if and only if $a\neq\frac{(\beta+1)(\overline{\beta}+1)}{\beta+\overline{\beta}}$. Hence, 
                    $$\left\lvert\{a : a \in \gf_q^* \text{ and } (\beta+\overline{\beta})a+(\beta+1)(\overline{\beta}+1)\neq0\}\right\rvert=q-2.$$
                \item When $\beta\in\gf_q$ and $\beta\neq1$, $(\beta+1)(\overline{\beta}+1)\neq0$. Then $(\beta + \overline{\beta}) a + (\beta + 1)(\overline{\beta} + 1) \neq 0$ holds for all $a \in \mathbb{F}_q^*$. Hence, 
                    $$\left\lvert\{a : a \in \gf_q^* \text{ and } (\beta+\overline{\beta})a+(\beta+1)(\overline{\beta}+1)\neq0\}\right\rvert=q-1.$$
            \end{itemize}
        \item For any $\beta\in\gf_{q^2}\setminus\{1\}$, we have
            \begin{align*}
                &\sum\limits_{\substack{a_1,a_2\in\gf_q\\\beta\overline{\beta}+(\beta+\overline{\beta})(a_1+a_2+1)+1\neq0}}(-1)^{\tr^s_1\left(\frac{a_2^2(a_1+a_2+\beta+1)(a_1+a_2+\overline{\beta}+1)}{(\beta\overline{\beta}+(\beta+\overline{\beta})(a_1+a_2+1)+1)^2}\right)}\\
                =&\sum\limits_{\substack{a_1,a_2\in\gf_q\\\beta\overline{\beta}+(\beta+\overline{\beta})(a_1+a_2+1)+1\neq0}}(-1)^{\tr^s_1\left(\frac{a_2^2((a_1+a_2)^2+(\beta\overline{\beta}+(\beta+\overline{\beta})(a_1+a_2+1)+1))}{(\beta\overline{\beta}+(\beta+\overline{\beta})(a_1+a_2+1)+1)^2}\right)}\\
                =&\sum\limits_{\substack{a_1,a_2\in\gf_q\\\beta\overline{\beta}+(\beta+\overline{\beta})(a_1+a_2+1)+1\neq0}}(-1)^{\tr^s_1\left(\frac{a_2^2(a_1+a_2)^2}{(\beta\overline{\beta}+(\beta+\overline{\beta})(a_1+a_2+1)+1)^2}\right)+\tr^s_1\left(\frac{a_2^2(\beta\overline{\beta}+(\beta+\overline{\beta})(a_1+a_2+1)+1)}{(\beta\overline{\beta}+(\beta+\overline{\beta})(a_1+a_2+1)+1)^2}\right)}\\
                =&\sum\limits_{\substack{a_1,a_2\in\gf_q\\\beta\overline{\beta}+(\beta+\overline{\beta})(a_1+a_2+1)+1\neq0}}(-1)^{\tr^s_1\left(\frac{a_2(a_1+a_2)}{\beta\overline{\beta}+(\beta+\overline{\beta})(a_1+a_2+1)+1}\right)+\tr^s_1\left(\frac{a_2^2}{\beta\overline{\beta}+(\beta+\overline{\beta})(a_1+a_2+1)+1}\right)}\\
                =&\sum\limits_{\substack{a_1,a_2\in\gf_q\\\beta\overline{\beta}+(\beta+\overline{\beta})(a_1+a_2+1)+1\neq0}}(-1)^{\tr^s_1\left(\frac{a_1a_2}{\beta\overline{\beta}+(\beta+\overline{\beta})(a_1+a_2+1)+1}\right)}.\\
            \end{align*}
            When $\beta\in\gf_q\setminus\{1\}$, we have $\beta+\overline{\beta}=0$.
            \begin{align*}
                &\sum\limits_{\substack{a_1,a_2\in\gf_q\\\beta\overline{\beta}+(\beta+\overline{\beta})(a_1+a_2+1)+1\neq0}}(-1)^{\tr^s_1\left(\frac{a_1a_2}{\beta\overline{\beta}+(\beta+\overline{\beta})(a_1+a_2+1)+1}\right)}\\
                =&\sum\limits_{a_1,a_2\in\gf_q}(-1)^{\tr^s_1\left(\frac{a_1a_2}{(\beta+1)(\overline{\beta}+1)}\right)}\\
                =&q + \sum\limits_{a_1\in\gf_q^*}\sum\limits_{a_2\in\gf_q}(-1)^{\tr^s_1\left(\frac{a_1a_2}{(\beta+1)(\overline{\beta}+1)}\right)}\\
                =&q.
            \end{align*}
            When $\beta\not\in\gf_q$,
            \begin{align*}
                &\sum\limits_{\substack{a_1,a_2\in\gf_q\\\beta\overline{\beta}+(\beta+\overline{\beta})(a_1+a_2+1)+1\neq0}}(-1)^{\tr^s_1\left(\frac{a_1a_2}{\beta\overline{\beta}+(\beta+\overline{\beta})(a_1+a_2+1)+1}\right)}\\
                =&q-1 + \sum\limits_{a_1\in\gf_q^*}\sum\limits_{a_2\in\gf_q\setminus\{a_1+\frac{(\beta+1)(\overline{\beta}+1)}{(\beta+\overline{\beta})^2}\}}(-1)^{\tr^s_1\left(\frac{a_1a_2}{\beta\overline{\beta}+(\beta+\overline{\beta})(a_1+a_2+1)+1}\right)}\\
                =&\left[1+(-1)^{\tr^s_1\left(\frac{(\beta+1)(\overline{\beta}+1)^2}{(\beta+\overline{\beta})}\right)}\right](q-1) + \sum\limits_{\substack{a_1\in\gf_q^*\\a_1\neq\frac{(\beta+1)(\overline{\beta}+1)}{\beta+\overline{\beta}}}}\sum\limits_{\substack{a_2\in\gf_q\\a_2\neq a_1+\frac{(\beta+1)(\overline{\beta}+1)}{\beta\overline{\beta}}}}(-1)^{\tr^s_1\left(\frac{a_1a_2}{\beta\overline{\beta}+(\beta+\overline{\beta})(a_1+a_2+1)+1}\right)}.\\
            \end{align*}
            Note that for any $a_2\neq a_2^{'}\in\gf_q$, $\frac{a_1a_2}{\beta\overline{\beta}+(\beta+\overline{\beta})(a_1+a_2+1)+1}=\frac{a_1a_2^{'}}{\beta\overline{\beta}+(\beta+\overline{\beta})(a_1+a_2^{'}+1)+1}$ holds only when $a_1=\frac{(\beta+1)(\overline{\beta}+1)}{\beta+\overline{\beta}}$ or $a_1=0$. Consequently,
            \begin{align*}
                &\sum\limits_{\substack{a_1\in\gf_q^*\\a_1\neq\frac{(\beta+1)(\overline{\beta}+1)}{\beta+\overline{\beta}}}}\sum\limits_{\substack{a_2\in\gf_q\\a_2\neq a_1+\frac{(\beta+1)(\overline{\beta}+1)}{\beta\overline{\beta}}}}(-1)^{\tr^s_1\left(\frac{a_1a_2}{\beta\overline{\beta}+(\beta+\overline{\beta})(a_1+a_2+1)+1}\right)}\\
                =&\sum\limits_{\substack{a_1\in\gf_q^*\\a_1\neq\frac{(\beta+1)(\overline{\beta}+1)}{\beta+\overline{\beta}}}}\sum\limits_{a_3\in\gf_q\setminus\{\frac{a_1}{\beta+\overline{\beta}}\}}(-1)^{\tr^s_1\left(a_3\right)}\\
                =&\sum\limits_{\substack{a_1\in\gf_q^*\\a_1\neq\frac{(\beta+1)(\overline{\beta}+1)}{\beta+\overline{\beta}}}}\left[\sum_{a_3\in\gf_q}(-1)^{\tr_1^s(a_3)-(-1)^{\tr_1^s(\frac{a_1}{\beta+\overline{\beta}})}}\right]\\
                =&-\sum\limits_{\substack{a_1\in\gf_q^*\\a_1\neq\frac{(\beta+1)(\overline{\beta}+1)}{\beta+\overline{\beta}}}}(-1)^{\tr_1^s(\frac{a_1}{\beta+\overline{\beta}})}\\
                =&-\sum_{a_1\in\gf_q}(-1)^{\tr_1^s(\frac{a_1}{\beta+\overline{\beta}})}+1+(-1)^{\tr_1^s\left(\frac{(\beta+1)+(\overline{\beta}+1)}{(\beta+\overline{\beta})^2}\right)}\\
                =&1+(-1)^{\tr_1^s\left(\frac{(\beta+1)+(\overline{\beta}+1)}{(\beta+\overline{\beta})^2}\right)}.
            \end{align*}
            Note that by Lemma \ref{lem:values_for_addsum} 
            \begin{align*}
                \tr^s_1\left(\frac{(\beta+1)(\overline{\beta}+1)}{(\beta+\overline{\beta})^2}\right)=&\tr^s_1\left(\frac{\beta\overline{\beta}}{(\beta+\overline{\beta})^2}\right) + \tr^s_1\left(\frac{\beta+\overline{\beta}}{(\beta+\overline{\beta})^2}\right) + \tr^s_1\left(\frac{1}{(\beta+\overline{\beta})^2}\right)\\
                =&\tr^s_1\left(\frac{\beta\overline{\beta}}{(\beta+\overline{\beta})^2}\right)\\
                =&1.
            \end{align*}
            Combining the above, we obtain
                $$\sum\limits_{\substack{a_1,a_2\in\gf_q\\\beta\overline{\beta}+(\beta+\overline{\beta})(a_1+a_2+1)+1\neq0}}(-1)^{\tr^s_1\left(\frac{a_1a_2}{\beta\overline{\beta}+(\beta+\overline{\beta})(a_1+a_2+1)+1}\right)}=\left[1+(-1)^{\tr^s_1\left(\frac{(\beta+1)(\overline{\beta}+1)}{(\beta+\overline{\beta})^2}\right)}\right]q=0.$$
    \end{itemize}
\end{proof}

The following lemma is essential for computing $\rho\bigl(\mathcal{C}_{(q,q^m+1,3,0)}\bigr)$ for $m>1$.
\begin{lemma}\label{lem:addsum_for_CR}
    Let $p=2$, $q=2^s$ and $m\in\Z^{+}$. Let $\beta$ be an arbitrary element in $\gf_{q^m}$.
    \begin{itemize}
        \item $\lvert \{ (u_1,u_2) : u_1,u_2\in U_{q^m+1}, u_1+u_2+\beta\neq0 \} \rvert=\left\{
	            \begin{aligned}
                    &q^m(q^m+1),& &\mbox{if}~~\beta=0,\\
                    &q^{2m}+2q^m-1,& &\mbox{if}~~\tr^{sm}_1(\beta^{-1})=1,\\
		            &(q^m+1)^2,& &\mbox{if}~~\tr^{sm}_1(\beta^{-1})=0.
	            \end{aligned}\right.$
        \item $\lvert \{ (u_1,u_2,u_3) : u_1,u_2,u_3\in U_{q^m+1}, u_1+u_2+u_3\neq0 \} \rvert =\left\{
	            \begin{aligned}
                    &(q^m+1)^3,& &\mbox{if}~~sm~~\mbox{is even},\\
                    &q^{3m}+3q^{2m}+q^m-1,& &\mbox{if}~~sm~~\mbox{is odd}.
	            \end{aligned}\right.$
        \item $\sum\limits_{\substack{u\in U_{q^m+1}\\u+\beta\neq0}}(-1)^{\tr^{sm}_1\left(\frac{1}{(u+\beta)^{q^m+1}}\right)}\left\{
                \begin{aligned}
                    &=(-1)^{sm}\left(q^m+1\right),& &\mbox{if}~~\beta=0,\\
                    &=-q^m,& &\mbox{if}~~\beta=1,\\
		            &\leqslant 2+2\sqrt{q^m},& &\mbox{if}~~\beta\notin\{0,1\}.
	            \end{aligned}\right.$
        \item $\sum\limits_{\substack{u_1,u_2\in U_{q^m+1}\\u_1+u_2+\beta\neq0}}(-1)^{\tr^{sm}_1\left(\frac{1}{(u_1+u_2+\beta)^{q^m+1}}\right)}\left\{
	            \begin{aligned}
                    &=-q^m(q^m+1),& &\mbox{if}~~\beta=0,\\
                    &\leqslant q^m(3+2\sqrt{q^m})+1,& &\mbox{if}~~\beta=1,\\
                    &\leqslant(q^m+2)(1+2\sqrt{q^m})+1,& &\mbox{if}~~\beta\not\in\{0,1\}.\\
	            \end{aligned}\right.$
    \end{itemize}
\end{lemma}
\begin{proof} We prove the four equalities in turn, as detailed below.
    \begin{itemize}
        \item Note that $u_2+\beta\in U_{q^m+1}$ if and only if $(u_2+\beta)(u_2^{-1}+\beta)=1$, i.e.,
                \begin{equation}\label{eqt:addsum_for_CR_1}
                    \beta\left(u_2^2+\beta u_2+1\right)=0.
                \end{equation}
            When $\beta=0$, 
                $$\sum\limits_{\substack{u_1,u_2\in U_{q^m+1}\\u_1+u_2+\beta\neq0}}1=\sum\limits_{\substack{u_1,u_2\in U_{q^m+1}\\u_1\neq u_2}}1=q^m(q^m+1).$$
            When $\beta\neq0$, (\ref{eqt:addsum_for_CR_1}) can be reduced to
                \begin{equation}\label{eqt:addsum_for_CR_2}
                    u_2^2+\beta u_2+1=0.
                \end{equation}
                Based on Lemma \ref{lem:sol_in_unit_peq2} and the condition that $\beta\in\gf_{q^m}$, one can verify that (\ref{eqt:addsum_for_CR_2}) has two distinct solutions in $U_{q^m+1}$ if and only if 
                $$\tr^{sm}_1\left(\frac{1}{\beta^2}\right)=\tr^{sm}_1(\beta^{-1})=1.$$
            Hence, when $\tr^{sm}_1(\beta^{-1})=1$, there are two elements in $U_{q^m+1}$(denoted by $x_2^{(1)}$ and $x_2^{(2)}$) such that $x_2^{(i)}+\beta\in U_{q^m+1}(i=1,2)$. Thus, under this condition, 
            $$\sum\limits_{\substack{u_1,u_2\in U_{q^m+1}\\u_1+u_2+\beta\neq0}}1=\sum\limits_{u_2\in U_{q^m+1}}\sum\limits_{\substack{u_1\in U_{q^m+1}\\u_1\notin \{x_2^{(1)}+\beta,x_2^{(2)}+\beta\}}}1=2q^m+(q^m-1)(q^m+1).$$
            When $\tr^{sm}_1(\beta^{-1})=0$, no element $u$ in $U_{q^m+1}$ such that $u+\beta$ belongs to $U_{q^m+1}$. Thus,
            $$\sum\limits_{\substack{u_1,u_2\in U_{q^m+1}\\u_1+u_2+\beta\neq0}}1=\sum\limits_{u_1,u_2\in U_{q^m+1}}1=(q^m+1)^2.$$
        \item The verification follows a similar analysis as in the first item, so the proof is omitted.
        \item When $\beta=0$, 
                $$\sum\limits_{\substack{u\in U_{q^m+1}\\u+\beta\neq0}}(-1)^{\tr^{sm}_1\left(\frac{1}{(u+\beta)^{q^m+1}}\right)}=\sum\limits_{u\in U_{q^m+1}}(-1)^{\tr^{sm}_1\left(1\right)}=(-1)^{sm}\left(q^m+1\right).$$
            When $\beta=1$, 
                $$\sum\limits_{\substack{u\in U_{q^m+1}\\u+\beta\neq0}}(-1)^{\tr^{sm}_1\left(\frac{1}{(u+\beta)^{q^m+1}}\right)}=\sum\limits_{u\in U_{q^m+1}\setminus\{1\}}(-1)^{\tr^{sm}_1\left(\frac{1}{u+u^{-1}}\right)}=-q^m.$$
            When $\beta\notin\{0,1\}$, according to Lemma \ref{lem:addsum_uB_on_u}, we have
                \begin{align*}
                    \sum\limits_{\substack{u\in U_{q^m+1}\\u+\beta\neq0}}(-1)^{\tr^{sm}_1\left(\frac{1}{(u+\beta)^{q^m+1}}\right)}&=(-1)^{\tr^{sm}_1\left(\frac{1}{(1+\beta)^{q^m+1}}\right)}+\sum\limits_{u\in U_{q^m+1}\setminus\{1\}}(-1)^{\tr^{sm}_1\left(\frac{1}{(u+\beta)^{q^m+1}}\right)}\\
                    &=(-1)^{\tr^{sm}_1\left((\beta+1)^{-1}\right)}+\sum\limits_{u\in U_{q^m+1}\setminus\{1\}}(-1)^{\tr^{sm}_1\left(\frac{1}{\beta u+\beta u^{-1}+(\beta+1)^2}\right)}\\
                    &\leqslant 2+2\sqrt{q^m}.
                \end{align*}
        \item Note that
                \begin{align*}
                    &\sum\limits_{\substack{u_1,u_2\in U_{q^m+1}\\u_1+u_2+\beta\neq0}}(-1)^{\tr_1^{sm}\left(\frac{1}{(u_1+u_2+\beta)^{q^m+1}}\right)}\\
                    =&\sum\limits_{\substack{u_1,u_2\in U_{q^m+1}\\u_1+u_2+\beta\neq0}}(-1)^{\tr_1^{sm}\left(\frac{1}{(u_1^{-1}+u_2^{-1}+\beta)(u_1+u_2+\beta)}\right)}\\
                    =&\sum\limits_{u_1\in U_{q^m+1}}\sum\limits_{\substack{u_2\in U_{q^m+1}\\u_2\neq u_1+\beta}}(-1)^{\tr_1^{sm}\left(\frac{1}{(u_1^{-1}+\beta)u_2+(u_1+\beta)u_2^{-1}+\beta(u_1^{-1} + u_1)+\beta^2}\right)}.
                \end{align*}
                When $\beta=0$, 
                \begin{align*}
                    &\sum\limits_{u_1\in U_{q^m+1}}\sum\limits_{\substack{u_2\in U_{q^m+1}\\u_2\neq u_1+\beta}}(-1)^{\tr_1^{sm}\left(\frac{u_1u_2}{(\beta u_1+1)u_2^2+\beta(u_1^2+\beta u_1+1)u_2+(u_1^2+\beta u_1)}\right)}\\
                    =&\sum\limits_{u_1\in U_{q^m+1}}\sum\limits_{\substack{u_2\in U_{q^m+1}\\u_2\neq u_1}}(-1)^{\tr_1^{sm}\left(\frac{u_1u_2}{u_2^2+u_1^2}\right)}\\
                    =&-q^m(q^m+1).
                \end{align*}
                When $\beta=1$, by Lemma \ref{lem:addsum_uB_on_u},
                \begin{align*}
                    &\sum\limits_{u_1\in U_{q^m+1}}\sum\limits_{\substack{u_2\in U_{q^m+1}\\u_2\neq u_1+\beta}}(-1)^{\tr_1^{sm}\left(\frac{u_1u_2}{(\beta u_1+1)u_2^2+\beta(u_1^2+\beta u_1+1)u_2+(u_1^2+\beta u_1)}\right)}\\
                    =&(q^m+1)(-1)^{\tr_1^{sm}(1)}+\sum\limits_{u_1\in U_{q^m+1}\setminus\{1\}}\sum\limits_{\substack{u_2\in U_{q^m+1}\\u_2\neq u_1+1}}(-1)^{\tr_1^{sm}\left(\frac{1}{(u_1^{-1}+1)u_2+(u_1+1)u_2^{-1}+u_1+u_1^{-1}+1}\right)}\\
                    \leqslant&q^m(3+2\sqrt{q^m})+1.
                \end{align*}
                When $\beta\not\in\{0,1\}$, it follows from Lemma \ref{lem:addsum_uB_on_u} that
                \begin{align*}
                    &\sum\limits_{u_1\in U_{q^m+1}}\sum\limits_{\substack{u_2\in U_{q^m+1}\\u_2\neq u_1+\beta}}(-1)^{\tr_1^{sm}\left(\frac{1}{(u_1^{-1}+\beta)u_2+(u_1+\beta)u_2^{-1}+\beta(u_1^{-1} + u_1)+\beta^2}\right)}\\
                    =&\sum\limits_{u_1\in U_{q^m+1}}\sum\limits_{\substack{u_2\in U_{q^m+1}\\u_2\notin \{1,u_1+\beta\}}}(-1)^{\tr_1^{sm}\left(\frac{1}{(u_1^{-1}+\beta)u_2+(u_1+\beta)u_2^{-1}+\beta(u_1^{-1} + u_1)+\beta^2}\right)}\\
                    &+\sum\limits_{u_1\in U_{q^m+1}\setminus\{1\}}(-1)^{\tr_1^{sm}\left(\frac{1}{(\beta+1)(u_1^{-1} + u_1)+\beta^2}\right)}+(-1)^{\tr_1^{sm}\left(\beta^{-1}\right)}\\
                    &\leqslant (q^m+2)(1+2\sqrt{q^m})+1.
                \end{align*}
    \end{itemize}
    The proof of Lemma \ref{lem:addsum_for_CR} is complete.
\end{proof}

\subsection{Character sums over finite fields of odd characteristic}

The following two quadratic character sums will be critical in Subsection \ref{sec:CR_0}; we state them as lemmas below.

\begin{lemma}\label{lem:chsum_odd_roots}
    Let $\beta$ be an arbitrary element in $\gf_{q^{2m}}$ with $q$ odd. Define 
        $$R_{\beta}=\left\{u\in U_{q^m+1} : (\beta-u)^{q^m+1}\left((\beta-u)^{q^m+1}-1\right)=0\right\}.$$ 
    Then
    $$|R_{\beta}|=\left\{
	    \begin{aligned}
            &q^m+1,& &\mbox{if}~~\beta=0,\\
            &3,& &\mbox{if}~~\beta\in U_{q^m+1}~\mbox{and}~\chi_{q^m}(-3)=-1,\\
		    &2,& &\mbox{if}~~\beta\notin U_{q^m+1}~\mbox{and}~\chi_{q^m}(\overline{\beta}\beta(\overline{\beta}\beta-4))=-1; \mbox{or}~p=3~\mbox{and}~\beta\in U_{q^m+1},\\
            &1,& &\mbox{if}~~\beta\notin U_{q^m+1}~\mbox{and}~\overline{\beta}\beta=4;\mbox{or}~\chi_{q^m}(-3)=1~\mbox{and}~\beta\in U_{q^m+1},\\
            &0,& &\mbox{if}~~\beta\notin U_{q^m+1}~\mbox{and}~\chi_{q^m}(\overline{\beta}\beta(\overline{\beta}\beta-4))=1.
	    \end{aligned}\right.$$
\end{lemma}
\begin{proof}
   Since the case that $\beta=0$ is trivial, we henceforth consider that $\beta\in\gf_{q^{2m}}^*$. 
    Note that $(\beta-u)^{q^m+1}=0$ iff $u=\beta$, while $(\beta-u)^{q^m+1}-1=0$ iff
        \begin{equation}\label{eqt:R_u1_1}
            u^2-\beta u+\beta^{1-q^m}=0.
        \end{equation}
    The discriminant of (\ref{eqt:R_u1_1}) is $$\Delta_u=\beta^{1-q^m}(\beta^{q^m+1}-4)$$ and it is straightforward to verify that $\chi_{q^{2m}}(\Delta_u)=1$. According to Lemma \ref{lem:sol_in_unit_p}, the solutions to (\ref{eqt:R_u1_1}) in $U_{q^m+1}$ depends on parameters as follows:
        \begin{itemize}
            \item Two distinct solutions exist if and only if $\chi_{q^{m}}(\Delta_u)=-1$;
            \item Exactly one solution exists if and only if $\Delta_u=0$;
            \item No solution exists if and only if $\chi_{q^{m}}(\Delta_u)=1$.
        \end{itemize}
    Given that $\beta$ is not a root of $(\beta-u)^{q^m+1}-1=0$, the value of $|R_{\beta}|$ follows directly.
\end{proof}

\begin{lemma}\label{lem:chsum_odd_1}
    Let $\beta$ be an arbitrary element in $\gf_{q^{2m}}$ with $q$ odd. Define $$s_{\beta}=\sum\limits_{u\in U_{q^m+1}\setminus R_{\beta}}\chi_{q^m}\left((\beta-u)^{-q^m-1}\left((\beta-u)^{q^m+1}-1\right)\right).$$ Then
        $$s_{\beta}=\left\{
	        \begin{aligned}
                &0,& &\mbox{if}~~\beta=0,\\
		        &-q^m+1,& &\mbox{if}~~p=3,\beta\in U_{q^m+1},\\
                &1,& &\mbox{if}~~p>3,\beta\in U_{q^m+1},\\
                &\chi_{q^m}(-2),& & \mbox{if}~~p>3,\beta\overline{\beta}=4,\beta\not\in U_{q^m+1}.\\
	        \end{aligned}\right.$$
    Otherwise, when $\beta\in\gf_{q^{2m}}^*\setminus U_{q^m+1}$, $$\left|s_{\beta}\right|\leqslant1+3\sqrt{q^m}.$$
\end{lemma}
\begin{proof}
    For $\beta\in U_{q^m+1}$, we have
    \begin{align*}
        s_{\beta}&=\sum\limits_{u\in U_{q^m+1}\setminus R_{\beta}}\chi_{q^m}\left((\beta-u)^{-q^m-1}\left((\beta-u)^{q^m+1}-1\right)\right)\\
                &=\sum\limits_{v\in U_{q^m+1}\setminus R_{1}}\chi_{q^m}\left((1-v)^{-q^m-1}\left((1-v)^{q^m+1}-1\right)\right)\\
                &=\sum\limits_{v\in U_{q^m+1}\setminus R_{1}}\chi_{q^m}\left((1-v)^{q^m+1}\left((1-v)^{q^m+1}-1\right)\right)\\
                &=\sum\limits_{v\in U_{q^m+1}}\chi_{q^m}\left((1-v)^{q^m+1}\left((1-v)^{q^m+1}-1\right)\right)\\
                &=\sum\limits_{v\in U_{q^m+1}\setminus\{1\}}\chi_{q^m}\left((v+v^{-1}-2)(v+v^{-1}-1)\right)\\
                &=\sum\limits_{\gamma\in \gf_{q^m}}\chi_{q^m}\left[\left(\frac{\gamma+\overline{A}}{\gamma+A}+\frac{\gamma+A}{\gamma+\overline{A}}-2\right)\left(\frac{\gamma+\overline{A}}{\gamma+A}+\frac{\gamma+A}{\gamma+\overline{A}}-1\right)\right]\\
                &=\sum\limits_{\gamma\in\gf_{q^m}}\chi_{q^m}\left[\left(\overline{A}-A\right)^2\left(\gamma^2+(\overline{A}+A)\gamma+(\overline{A}^2-\overline{A}A+A^2)\right)\right]\\
                &=(-1)\sum\limits_{\gamma\in\gf_{q^m}}\chi_{q^m}\left(\gamma^2+(\overline{A}+A)\gamma+(\overline{A}^2-\overline{A}A+A^2)\right).
    \end{align*}
    The second equality follows from the substitution $u\mapsto \beta v$. Note that the discriminant of the quadratic polynomial $\gamma^2+(\overline{A}+A)\gamma+(\overline{A}^2-\overline{A}A+A^2)$ is $-3(\overline{A}-A)^2$, then according to Lemma \ref{lem:chsum_deg2},
        $$s_{\beta}=\left\{
	        \begin{aligned}
                &-q^m+1,& &\mbox{if}~~p=3\\
		        &1,& &\mbox{if}~~p>3.
	        \end{aligned}\right.$$
   For $\beta\notin U_{q^m+1}$, we have
        \begin{align*}
            s_{\beta}&=\sum\limits_{u\in U_{q^m+1}\setminus R_{\beta}}\chi_{q^m}\left((\beta-u)^{q^m+1}\left((\beta-u)^{q^m+1}-1\right)\right)\\
                    &=\sum\limits_{u\in U_{q^m+1}}\chi_{q^m}\left((\beta-u)^{q^m+1}\left((\beta-u)^{q^m+1}-1\right)\right)\\
                    &=\sum\limits_{u\in U_{q^m+1}}\chi_{q^m}\left(\left(\overline{\beta}\beta-\overline{\beta}u-\beta u^{-1}+1\right)\left(\overline{\beta}\beta-\overline{\beta}u-\beta u^{-1}\right)\right)\\
                    &=\chi_{q^m}\left(\left(\overline{\beta}\beta-\overline{\beta}-\beta+1\right)\left(\overline{\beta}\beta-\overline{\beta}-\beta\right)\right)+\\
                    &~\sum\limits_{\gamma\in \gf_{q^m}}\chi_{q^m}\left(\left(\overline{\beta}\beta-\overline{\beta}\frac{\gamma+\overline{A}}{\gamma+A}-\beta \frac{\gamma+A}{\gamma+\overline{A}}+1\right)\left(\overline{\beta}\beta-\overline{\beta}\frac{\gamma+\overline{A}}{\gamma+A}-\beta \frac{\gamma+A}{\gamma+\overline{A}}\right)\right)\\
                    &=\chi_{q^m}\left(\left(\overline{\beta}\beta-\overline{\beta}-\beta+1\right)\left(\overline{\beta}\beta-\overline{\beta}-\beta\right)\right)+\\
                    &~\sum\limits_{\gamma\in\gf_{q^m}}\chi_{q^m}\left[\left(\left(\overline{\beta}\beta - \overline{\beta} - \beta\right)\gamma^{2} + \left(\overline{\beta}\beta(A+\overline{A}) - 2\overline{\beta}\overline{A} - 2\beta A\right)\gamma + \left(\overline{\beta}\beta \overline{A}A - \overline{\beta}\overline{A}^{2} - \beta A^{2}\right)\right)\cdot\right.\\
                    &~~~~~~~~~~~~~~~~~~
                    \left. \left((\beta-1)\gamma+\beta A-\overline{A}\right)\left((\overline{\beta}-1)\gamma+\overline{\beta}\overline{A}-A\right)\right].
        \end{align*}
        Set $g_{\beta}(\gamma)=\left(\overline{\beta}\beta - \overline{\beta} - \beta\right)\gamma^{2} + \left(\overline{\beta}\beta(A+\overline{A}) - 2\overline{\beta}\overline{A} - 2\beta A\right)\gamma + \left(\overline{\beta}\beta \overline{A}A - \overline{\beta}\overline{A}^{2} - \beta A^{2}\right)$. The discriminant of $g_{\beta}(\gamma)$ is $\Delta(g_{\beta})=\beta\overline{\beta}(\beta\overline{\beta}-4)(\overline{A}-A)^2$. For the case $p>3$ and $\beta=2$, we have
            \begin{align*}
            s_{2}&=\sum\limits_{\gamma\in\gf_{q^m}}\chi_{q^m}\left[-2\left(\overline{A}-A\right)^2\left(\gamma+2A-\overline{A}\right)\left(\gamma+2\overline{A}-A\right)\right]\\  
            &=\chi_{q^m}(-2).
        \end{align*}
        The last identity follows from the fact that $\chi_{q^m}\left((\overline{A}-A)^2\right)=-1$ and the discriminant of $\left(\gamma+2A-\overline{A}\right)\left(\gamma+2\overline{A}-A\right)$ is $9(A-\overline{A})^2$. 
        For the case $p>3$, $\beta\overline{\beta}=4$, and $\beta\neq2$, we obtain
        \begin{align*}
            s_{\beta}&=\chi_{q^m}\left((\overline{\beta}+\beta-4)(\overline{\beta}+\beta-5)\right) + \\
                    &~\sum\limits_{\gamma\in\gf_{q^m}}\chi_{q^m}\left[(4-\beta-\overline{\beta})\left(\gamma-\frac{2\overline{A}-A\beta}{\beta-2}\right)^2\left((\beta-1)\gamma+\beta A-\overline{A}\right)\left((\overline{\beta}-1)\gamma+\overline{\beta}\overline{A}-A\right)\right]\\  
                    &=\chi_{q^m}\left((\overline{\beta}+\beta-4)(\overline{\beta}+\beta-5)\right) + \sum\limits_{\gamma\in\gf_{q^m}\setminus\{\frac{2\overline{A}-A\beta}{\beta-2}\}}\chi_{q^m}\left[(4-\beta-\overline{\beta})\left((\beta-1)\gamma+\beta A-\overline{A}\right)\left((\overline{\beta}-1)\gamma+\overline{\beta}\overline{A}-A\right)\right]\\
                    &=\chi_{q^m}\left((\overline{\beta}+\beta-4)(\overline{\beta}+\beta-5)\right) + \chi_{q^m}(-2) + \\
                    &~\sum\limits_{\gamma\in\gf_{q^m}}\chi_{q^m}\left[(4-\beta-\overline{\beta})\left((\beta-1)\gamma+\beta A-\overline{A}\right)\left((\overline{\beta}-1)\gamma+\overline{\beta}\overline{A}-A\right)\right]\\
                    &=\chi_{q^m}\left((\overline{\beta}+\beta-4)(\overline{\beta}+\beta-5)\right) + \chi_{q^m}(-2) - \chi_{q^m}\left((4-\beta-\overline{\beta})(5-\beta-\overline{\beta})\right)\\
                    &=\chi_{q^m}(-2).
        \end{align*}
        In all remaining cases, when $\beta\overline{\beta}\neq4$, we claim that
            $$g_{\beta}(\gamma)\cdot\left((\beta-1)\gamma+\beta A-\overline{A}\right)\left((\overline{\beta}-1)\gamma+\overline{\beta}\overline{A}-A\right)$$
        is not a square in $\mathbb F_{q^m}[\gamma]$.
        Since neither $\frac{\overline{A}-A\beta}{\beta-1}$ and $\frac{A-\overline{A}\overline{\beta}}{\overline{\beta}-1}$ lies in $\gf_{q^m}$, then 
            $$\left((\beta-1)\gamma+\beta A-\overline{A}\right)\left((\overline{\beta}-1)\gamma+\overline{\beta}\overline{A}-A\right)$$
        is an irreducible quadratic polynomial over \(\mathbb F_{q^m}\). Additionally, $\frac{\overline{A}-A\beta}{\beta-1}$ is not a root of $g_{\beta}(\gamma)$ since 
            $$g_{\beta}(\frac{\overline{A}-A\beta}{\beta-1})=-\frac{\overline{\beta}(A-\overline{A})^2}{(\overline{\beta}-1)^2}\neq0.$$ 
        Consequently,
            $$g_{\beta}(\gamma)\cdot\left((\beta-1)\gamma+\beta A-\overline{A}\right)\left((\overline{\beta}-1)\gamma+\overline{\beta}\overline{A}-A\right)$$
        is a polynomial of $\gamma$ whose degree is at most $4$ and is not a square in $\mathbb F_{q^m}[\gamma]$. Applying Lemma \ref{lem:WeilBound_multi} then yields, 
            $$\left|s_{\beta}\right|\leqslant 1+3\sqrt{q^m}.$$ 
\end{proof}

\section{The covering radius of $\C_{(q,q+1,3,0)}$}\label{sec:CR_meq1}
This section is devoted to determining the covering radius of \(\mathcal C_{(q,q+1,3,0)}\), corresponding to the special case \(m=1\). We treat the cases of odd and even \(q\) separately. It is known from \cite{CaoCXW2025BCHq_1} that the code \(C_{(q,q+1,3,0)}\) has parameters $[q+1,q-2,4]$ when $q>2$. Moreover, its extension-field parity-check matrix can be written as
        $$H_{(q,q+1,3,0)}=\begin{pmatrix}
                1 & 1 & 1 & \cdots & 1 \\
                1 & \mu^{1} & \mu^{2} & \cdots & \mu^{q} 
            \end{pmatrix},$$
where \(\mu\) is a primitive \((q+1)\)-th root of unity and \(U_{q+1}=\langle \mu\rangle\).

The following lemma establishes an upper bound on $\rho\left(\C_{(q,q+1,3,0)}\right)$ when $q$ is odd.
\begin{lemma}\label{lem:meq1_qodd_ubound}
    Let $q>5$ be odd. For any $\beta_1\in\gf_q$ and $\beta_2\in\gf_{q^2}$, there exist $a_1,a_2\in\gf_q$ and $u_1,u_2\in U_{q+1}$ such that the following system holds:
    \begin{equation}\label{eqtsys:meq1_qodd_le2_1}
        \begin{cases}
            \begin{aligned}
                &a_1 + a_2 &=& \beta_1, \\
                &a_1u_1 + a_2u_2 &=& \beta_2.
            \end{aligned}
        \end{cases}
    \end{equation}
\end{lemma}
\begin{proof}
   Since the case $\beta_1 = \beta_2 = 0$ is trivial, we assume in what follows that $(\beta_1, \beta_2) \neq (0,0)$. The proof is divided into three cases: (i) $\beta_1 \neq 0$, $\beta_2 = 0$; (ii) $\beta_1^2=\beta_2\overline{\beta_2}$; and (iii) $\beta_2\neq0$, $\beta_1^2\neq\beta_2\overline{\beta_2}$.
   \begin{itemize}
    \item When $\beta_1\neq0$, $\beta_2=0$, one of the solutions to system \eqref{eqtsys:meq1_qodd_le2_1} is given by:
            $$\begin{cases}
                &a_1=a_2=\frac{\beta_1}{2},\\
                &u_1=-u_2, u_2\in U_{q+1}.
            \end{cases}$$
    \item When $\beta_1^2=\beta_2\overline{\beta_2}$, one of the solutions to system \eqref{eqtsys:meq1_qodd_le2_1} is given by:
            $$\begin{cases}
                &a_1\in\gf_q^*\setminus\{\beta_1\},\\
                &a_2 = \beta_1 - a_1, \\
                &u_1=u_2=\frac{\beta_1}{\overline{\beta_2}}.
            \end{cases}$$
    \item Next, we consider $\beta_2\neq0$ and $\beta_1^2\neq\beta_2\overline{\beta_2}$. From the first equation of \eqref{eqtsys:meq1_qodd_le2_1}, we have $a_2 = \beta_1 - a_1$. Substituting this into the second equation yields
        $$a_1u_1 + (\beta_1-a_1)u_2 = \beta_2.$$
        Assuming $a_1 \neq 0$, we obtain
            $$u_1=a_1^{-1}[\beta_2-(\beta_1-a_1)u_2].$$
        This $u_1$ lies in $U_{q+1}$ if and only if
            $$a_1^{-1}(\beta_2-(\beta_1-a_1)u_2)\cdot\overline{\left(a_1^{-1}(\beta_2-(\beta_1-a_1)u_2)\right)}=1,$$
        which is equivalent to
            $$(\beta_1-a_1)\overline{\beta_2}u_2^2 - (\beta_2\overline{\beta_2}+\beta_1^2-2a_1\beta_1)u_2 + (\beta_1-a_1)\beta_2 = 0.$$
        When $a_1\neq\beta_1$ and $\beta_2\overline{\beta_2}+\beta_1^2-2a_1\beta_1\neq0$, this reduces to
        \begin{equation}\label{eqt:meq1_qodd_le2_1}
            u_2^2 - \frac{\beta_2\overline{\beta_2}+\beta_1^2-2a_1\beta_1}{\overline{\beta_2}(\beta_1-a_1)}u_2 + \frac{\beta_2}{\overline{\beta_2}} = 0.
        \end{equation}
        A direct computation shows that
            $$\left(\frac{\beta_2\overline{\beta_2}+\beta_1^2-2a_1\beta_1}{\overline{\beta_2}(\beta_1-a_1)}\right)^2-4\frac{\beta_2}{\overline{\beta_2}}=\frac{\left(\beta_2\overline{\beta_2}+\beta_1^2-2a_1\beta_1\right)^2-4\beta_2\overline{\beta_2}(\beta_1-a_1)^2}{\overline{\beta_2}^2(\beta_1-a_1)^2}$$
        is square in $\gf_{q^2}^*$, and 
            $$\frac{\beta_2}{\overline{\beta_2}}=\frac{\frac{\beta_2\overline{\beta_2}+\beta_1^2-2a_1\beta_1}{\overline{\beta_2}(\beta_1-a_1)}}{\overline{\left(\frac{\beta_2\overline{\beta_2}+\beta_1^2-2a_1\beta_1}{\overline{\beta_2}(\beta_1-a_1)}\right)}}.$$
        Consequently, by Lemma \ref{lem:sol_in_unit_p}, \eqref{eqt:meq1_qodd_le2_1} has solutions in $U_{q+1}$ if and only if there exists $a_1\in\gf_q^*\setminus\{\beta_1\}$ such that 
             $$\theta+\overline{\theta}=0,\text{ where } \theta^2=\frac{(\beta_2\overline{\beta_2}+\beta_1^2-2a_1\beta_1)^2-4\beta_2\overline{\beta_2}(\beta_1-a_1)^2}{(\beta_2\overline{\beta_2}+\beta_1^2-2a_1\beta_1)^2}.$$
        Suppose that $\theta\neq0$, then it suffices to prove that there exists $a_1\in\gf_q^*\setminus\{\beta_1\}$ satisfying $\beta_2\overline{\beta_2}+\beta_1^2-2a_1\beta_1\neq0$ such that $\theta^{q-1}=-1$, that is, 
            $$\chi_q\left(\frac{(\beta_2\overline{\beta_2}+\beta_1^2-2a_1\beta_1)^2-4\beta_2\overline{\beta_2}(\beta_1-a_1)^2}{(\beta_2\overline{\beta_2}+\beta_1^2-2a_1\beta_1)^2}\right)=-1.$$
        Applying Lemma \ref{lem:chsum_deg2}, we compute
        \begin{align*}
            &\sum\limits_{\substack{a_1\in\gf_q^*\setminus\{\beta_1\}\\\beta_2\overline{\beta_2}+\beta_1^2-2a_1\beta_1\neq0\\(\beta_2\overline{\beta_2}+\beta_1^2-2a_1\beta_1)^2-4\beta_2\overline{\beta_2}(\beta_1-a_1)^2\neq0}}\chi_q\left(\frac{(\beta_2\overline{\beta_2}+\beta_1^2-2a_1\beta_1)^2-4\beta_2\overline{\beta_2}(\beta_1-a_1)^2}{(\beta_2\overline{\beta_2}+\beta_1^2-2a_1\beta_1)^2}\right) \\
            = & \sum\limits_{\substack{a_1\in\gf_q^*\setminus\{\beta_1\}\\\beta_2\overline{\beta_2}+\beta_1^2-2a_1\beta_1\neq0\\(\beta_2\overline{\beta_2}+\beta_1^2-2a_1\beta_1)^2-4\beta_2\overline{\beta_2}(\beta_1-a_1)^2\neq0}}\chi_q\left(((\beta_2\overline{\beta_2}+\beta_1^2-2a_1\beta_1)^2-4\beta_2\overline{\beta_2}(\beta_1-a_1)^2)(\beta_2\overline{\beta_2}+\beta_1^2-2a_1\beta_1)^2\right) \\
            = & \sum\limits_{\substack{a_1\in\gf_q^*\setminus\{\beta_1\}\\\beta_2\overline{\beta_2}+\beta_1^2-2a_1\beta_1\neq0}}\chi_q\left((\beta_2\overline{\beta_2}+\beta_1^2-2a_1\beta_1)^2-4\beta_2\overline{\beta_2}(\beta_1-a_1)^2\right). \\
        \end{align*}
        \begin{itemize}
            \item When $\beta_1=0$, 
                \begin{align*}
                    &\sum\limits_{\substack{a_1\in\gf_q^*\setminus\{\beta_1\}\\\beta_2\overline{\beta_2}+\beta_1^2-2a_1\beta_1\neq0}}\chi_q\left((\beta_2\overline{\beta_2}+\beta_1^2-2a_1\beta_1)^2-4\beta_2\overline{\beta_2}(\beta_1-a_1)^2\right)\\
                    =&\sum\limits_{a_1\in\gf_q^*}\chi_q\left(-4\beta_2\overline{\beta_2}a_1^2+\beta_2^2\overline{\beta_2}^2\right)\\
                    =&\sum\limits_{a_1\in\gf_q}\chi_q\left(-4\beta_2\overline{\beta_2}a_1^2+\beta_2^2\overline{\beta_2}^2\right) - 1\\
                    =&-\chi_q(-\beta_2\overline{\beta_2}) - 1.
                \end{align*}
            \item When $\beta_1\neq0$ and $\beta_1^2+\beta_2\overline{\beta_2} = 0$,
                \begin{align*}
                    &\sum\limits_{\substack{a_1\in\gf_q^*\setminus\{\beta_1\}\\\beta_2\overline{\beta_2}+\beta_1^2-2a_1\beta_1\neq0}}\chi_q\left((\beta_2\overline{\beta_2}+\beta_1^2-2a_1\beta_1)^2-4\beta_2\overline{\beta_2}(\beta_1-a_1)^2\right)\\
                    =&\sum\limits_{a_1\in\gf_q}\chi_q\left(4\beta_1^2 (2a_1^2 - 2\beta_1 a_1 + \beta_1^2)\right) - 2\chi_q(4\beta_1^4)\\
                    =&-\chi_q(2) - 2.
                \end{align*}
            \item When $\beta_1\neq0$ and $\beta_1^2+\beta_2\overline{\beta_2} \neq 0$,
                \begin{align*}
                    &\sum\limits_{\substack{a_1\in\gf_q^*\setminus\{\beta_1\}\\\beta_2\overline{\beta_2}+\beta_1^2-2a_1\beta_1\neq0}}\chi_q\left((\beta_2\overline{\beta_2}+\beta_1^2-2a_1\beta_1)^2-4\beta_2\overline{\beta_2}(\beta_1-a_1)^2\right)\\
                    =&\sum\limits_{a_1\in\gf_q}\chi_q\left(4(\beta_1^2 - \beta_2 \overline{\beta_2}) a_1^2 - 4\beta_1(\beta_1^2 - \beta_2 \overline{\beta_2}) a_1 + (\beta_1^2 - \beta_2 \overline{\beta_2})^2\right) - 2\chi_q((\beta_2\overline{\beta_2} - \beta_1^2)^2)\\
                    & - \chi_q(-\beta_2\overline{\beta_2})\\
                    =&-\chi_q(\beta_1^2 - \beta_2 \overline{\beta_2}) - \chi_q(-\beta_2\overline{\beta_2}) - 2.
                \end{align*}
                The last equality follows from the fact that the discriminant of the quadratic (in $a_1$)
                    $$4(\beta_1^2 - \beta_2 \overline{\beta_2}) a_1^2 - 4\beta_1(\beta_1^2 - \beta_2 \overline{\beta_2}) a_1 + (\beta_1^2 - \beta_2 \overline{\beta_2})^2$$
                is $16 \beta_2 \overline{\beta_2} (\beta_1^2 - \beta_2 \overline{\beta_2})^2 \neq 0$.
        \end{itemize}
        Therefore,
            $$\sum\limits_{\substack{a_1\in\gf_q^*\setminus\{\beta_1\}\\\beta_2\overline{\beta_2}+\beta_1^2-2a_1\beta_1\neq0\\(\beta_2\overline{\beta_2}+\beta_1^2-2a_1\beta_1)^2-4\beta_2\overline{\beta_2}(\beta_1-a_1)^2\neq0}}\frac{1-\chi_q\left(\frac{(\beta_2\overline{\beta_2}+\beta_1^2-2a_1\beta_1)^2-4\beta_2\overline{\beta_2}(\beta_1-a_1)^2}{(\beta_2\overline{\beta_2}+\beta_1^2-2a_1\beta_1)^2}\right)}{2}\geqslant\frac{1}{2}(q-5).$$
        Thus, for any $q>5$, there exists $a_1\in\gf_q^*\setminus\{\beta_1,\frac{\beta_2\overline{\beta_2}+\beta_1^2}{2\beta_1}\}$ such that \eqref{eqt:meq1_qodd_le2_1} possesses solutions in $U_{q+1}$. In short, one of the solutions to system \eqref{eqtsys:meq1_qodd_le2_1} is given by:
            $$\begin{cases}
                &a_1\in\gf_q^*\setminus\{\beta_1,\frac{\beta_2\overline{\beta_2}+\beta_1^2}{2\beta_1}\} \text{ satisfying } \chi_q\left(\frac{(\beta_2\overline{\beta_2}+\beta_1^2-2a_1\beta_1)^2-4\beta_2\overline{\beta_2}(\beta_1-a_1)^2}{(\beta_2\overline{\beta_2}+\beta_1^2-2a_1\beta_1)^2}\right)=-1,\\
                &a_2 = \beta_1 - a_1,\\
                &u_2\in U_{q+1} \text{ is a solution of } u_2^2 - \frac{\beta_2\overline{\beta_2}+\beta_1^2-2a_1\beta_1}{\overline{\beta_2}(\beta_1-a_1)}u_2 + \frac{\beta_2}{\overline{\beta_2}} = 0,\\
                &u_1=a_1^{-1}[\beta_2-(\beta_1-a_1)u_2].
            \end{cases}$$
   \end{itemize}
   Having treated all three cases, the proof is complete.
\end{proof}

The covering radius of $\left(\C_{(q,q+1,3,0)}\right)$ for odd $q$ is given in the following theorem.
\begin{theorem}\label{thm:meq1_qodd}
    When $q$ is odd, it holds that $\rho\left(\C_{(q,q+1,3,0)}\right)=2$.
\end{theorem}
\begin{proof}
    Recall that the parameters of $\mathcal{C}_{(q,q+1,3,0)}$ are $[q+1, q-2, 4]$. A direct verification shows that this code is not perfect; consequently, $\rho\bigl(\mathcal{C}_{(q,q+1,3,0)}\bigr) \geqslant 2$. On the other hand, by Lemma~\ref{lem:CR_PCmatrix} and Lemma \ref{lem:meq1_qodd_ubound} we have $\rho\bigl(\mathcal{C}_{(q,q+1,3,0)}\bigr) \le 2$ whenever $q > 5$. For the two exceptional cases $q=3$ and $q=5$, a direct computation in MAGMA using the intrinsic function \texttt{CoveringRadius} confirms that $\rho\bigl(\mathcal{C}_{(3,4,3,0)}\bigr) = 2$ and $\rho\bigl(\mathcal{C}_{(5,6,3,0)}\bigr) = 2$, respectively. Thus, in all cases, $\rho\bigl(\mathcal{C}_{(q,q+1,3,0)}\bigr) = 2$.
\end{proof}
\begin{remark}
    When $q$ is odd, the BCH code $\mathcal{C}_{(q,q+1,3,0)}$ is shown to be both maximum distance separable (MDS) and quasi-perfect.
\end{remark}

The following lemma provides a lower bound on $\rho\left(\C_{(q,q+1,3,0)}\right)$ when $q$ is even.
\begin{lemma}\label{lem:meq1_qeven_lbound}
    When $q$ is even, there exist $\beta_1 \in\gf_q,\beta_2 \in\gf_{q^{2}}$ such that for all $a_1,a_2\in\gf_q$ and $u_1,u_2\in U_{q+1}$, the following system has no solution.
    \begin{equation}\label{eqtsys:meq1_qodd_ge3_1}
        \begin{cases}
            \begin{aligned}
                &a_1 + a_2 &=& \beta_1, \\
                &a_1u_1 + a_2u_2 &=& \beta_2.
            \end{aligned}
        \end{cases}
    \end{equation}
\end{lemma}
\begin{proof}
    The proof proceeds by setting $\beta_1 = 1$ and $\beta_2 = 0$. System \eqref{eqtsys:meq1_qodd_ge3_1} then reduces to
        $$a_1u_1 + (1+a_1)u_2 = 0.$$
    If $a_1=0$, then $(1+a_1)u_2 = 0$ implies $a_1=1$ (since $u_2 \neq 0$), which contradicts $a_1 = 0$. Hence, $a_1 \neq 0$. Solving for $u_1$ yields
        $$u_1=\frac{1+a_1}{a_1}u_2.$$ 
    The condition $u_1 \in U_{q+1}$ requires
        $$\frac{1+a_1}{a_1}u_2\cdot\frac{1+a_1}{a_1}\overline{u_2}=\left(\frac{1+a_1}{a_1}\right)^2=1,$$
    which forces $1 + a_1 = a_1$, an impossibility. This completes the proof.
\end{proof}

The following lemma provides an upper bound on $\rho\left(\C_{(q,q+1,3,0)}\right)$ when $q$ is even. For \(q=2\), the code \(\mathcal{C}_{(2,3,3,0)}\) is the zero code. Hence, this degenerate case will be excluded from the discussion of covering radii.

\begin{lemma}\label{lem:meq1_qeven_ubound}
    When $q>2$ is even, for any $\beta_1\in\gf_q$ and $\beta_2\in\gf_{q^2}$, there exist $a_1,a_2,a_3\in\gf_q$ and $u_1,u_2,u_3\in U_{q+1}$ such that the following system holds:
    \begin{equation}\label{eqtsys:meq1_qeven_le3_1}
        \begin{cases}
            \begin{aligned}
                &a_1 + a_2 + a_3 &=& \beta_1, \\
                &a_1u_1 + a_2u_2 + a_3u_3&=& \beta_2.
            \end{aligned}
        \end{cases}
    \end{equation}
\end{lemma}
\begin{proof}
    Since the case $\beta_1 = \beta_2 = 0$ is trivial, we assume in what follows that $(\beta_1, \beta_2) \neq (0,0)$.
    The proof proceeds by considering $\beta_1=0$ and $\beta_1\neq0$, respectively.
    \begin{itemize}
        \item When $\beta_1=0$, system \eqref{eqtsys:meq1_qeven_le3_1} reduces to
                \begin{equation}\label{eqtsys:meq1_qeven_le3_2}
                    \begin{cases}
                        \begin{aligned}
                            &a_1 + a_2 + a_3 &=& 0, \\
                            &a_1u_1 + a_2u_2 + a_3u_3&=& \beta_2.
                        \end{aligned}
                    \end{cases}
                \end{equation}
            Setting $a_3 = 0$ and $a_1 = a_2$, we obtain
                $$a_1(u_1+u_2)=\beta_2.$$
            Since $\beta_2 \neq 0$, we have $a_1 \neq 0$ and $u_1 \neq u_2$. Solving for $u_1$ yields
                $$u_1 = u_2 + a_1^{-1}\beta_2.$$
            The condition $u_1\in U_{q+1}$ requires $(u_2 + a_1^{-1}\beta_2)(\overline{u_2 + a_1^{-1}\beta_2})=1,$ that is,
            \begin{equation}\label{eqt:meq1_qeven_le3_1}
                u_2^2 + \frac{\beta_2}{a_1}u_2 + \frac{\beta_2}{\overline{\beta_2}} = 0.
            \end{equation}
            A direct computation shows that
                $$\tr_1^{2s}\left(\frac{\beta_2}{\overline{\beta_2}}/\frac{\beta_2^2}{a_1^2}\right)=\tr_1^{2s}\left(\frac{a_1^2}{\beta_2\overline{\beta_2}}\right)=0, \text{ and } \frac{\beta_2}{\overline{\beta_2}}=\frac{\frac{\beta_2}{a_1}}{\overline{\left(\frac{\beta_2}{a_1}\right)}}.$$
            By Lemma \ref{lem:sol_in_unit_peq2}, \eqref{eqt:meq1_qeven_le3_1} has solutions in $U_{q+1}$ if and only if there exists $a_1\in\gf_q^*$ such that
                $$\tr_1^{s}\left(\frac{a_1^2}{\beta_2\overline{\beta_2}}\right)=1.$$
            Let $\beta_{21} \in \mathbb{F}_q^*$ be an element satisfying $\beta_{21}^2 = \beta_2 \overline{\beta_2}$. Then
                $$\sum\limits_{a_1\in\gf_q^*}(-1)^{\tr_1^{s}\left(\frac{a_1^2}{\beta_2\overline{\beta_2}}\right)} = \sum\limits_{a_1\in\gf_q}(-1)^{\tr_1^{s}\left(\frac{a_1^2}{\beta_2\overline{\beta_2}}\right)} - 1 = \sum\limits_{a_1\in\gf_q}(-1)^{\tr_1^{s}\left(\frac{a_1}{\beta_{21}}\right)} - 1 = -1.$$
            Consequently,
                $$\sum\limits_{a_1\in\gf_q^*}\frac{1-(-1)^{\tr_1^{s}\left(\frac{a_1^2}{\beta_2\overline{\beta_2}}\right)}}{2} = \frac{1}{2}\left[q-1-\sum\limits_{a_1\in\gf_q^*}(-1)^{\tr_1^{s}\left(\frac{a_1^2}{\beta_2\overline{\beta_2}}\right)}\right] = \frac{q}{2}.$$
            Thus, whenever $q>2$, there exists $a_1\in\gf_q^*$ such that \eqref{eqt:meq1_qeven_le3_1} possesses solutions in $U_{q+1}$. In this case, a solution to system \eqref{eqtsys:meq1_qeven_le3_2} is given by
            $$\begin{cases}
                &a_1\in\gf_q^* \text{ satisfying } \tr_1^{s}\left(\frac{a_1^2}{\beta_2\overline{\beta_2}}\right)=1,\\
                &a_2=a_1,\\
                &a_3=0,\\
                &u_1 = u_2 + a_1^{-1}\beta_2,\\
                &u_2 \text{ is a solution of } u_2^2 + \frac{\beta_2}{a_1}u_1 + \frac{\beta_2}{\overline{\beta_2}} = 0,\\
                &u_3\in U_{q+1}.
            \end{cases}$$
        \item When $\beta_1\neq0$, system \eqref{eqtsys:meq1_qeven_le3_1} reduces to 
                \begin{equation}\label{eqtsys:meq1_qeven_le3_3}
                    \begin{cases}
                        \begin{aligned}
                            &\frac{a_1}{\beta_1} + \frac{a_2}{\beta_1} + \frac{a_3}{\beta_1} &=& 1, \\
                            &\frac{a_1}{\beta_1}u_1 + \frac{a_2}{\beta_1}u_2 + \frac{a_3}{\beta_1}u_3 &=& \beta_2\beta_1^{-1}.
                        \end{aligned}
                    \end{cases}
                \end{equation}
                Hence, it suffices to show that for any $\beta_2 \in \mathbb{F}_{q^2}$, there exist $a_1, a_2, a_3 \in \mathbb{F}_q$ and $u_1, u_2, u_3 \in U_{q+1}$ satisfying
                \begin{equation}\label{eqtsys:meq1_qeven_le3_4}
                    \begin{cases}
                        \begin{aligned}
                            &a_1 + a_2 + a_3 &=& 1, \\
                            &a_1u_1 + a_2u_2 + a_3u_3 &=& \beta_2.
                        \end{aligned}
                    \end{cases}
                \end{equation}
                When $\beta_2=1$, observe that $(a_1,a_2,a_3,u_1,u_2,u_3)=(1,0,0,1,1,1)$ is a solution to system \eqref{eqtsys:meq1_qeven_le3_4}. When $\beta_2\neq1$,  the proof proceeds by setting $u_3=1$. From the first equation of system \eqref{eqtsys:meq1_qeven_le3_3}, we obtain 
                    $$a_3 = 1 + a_1 + a_2.$$
                Substituting this into the second equation yields
                    $$a_1u_1 + a_2u_2 + a_1 + a_2 + 1 = \beta_2.$$
                Assuming $a_1 \neq 0$, we solve for $u_1$ and get
                    $$u_1=a_1^{-1}(\beta_2 + a_2u_2 + a_1 + a_2 + 1).$$
                The condition $u_1\in U_{q+1}$ requires 
                    $$a_1^{-1}(\beta_2 + a_2u_2 + a_1 + a_2 + 1)\cdot\overline{(a_1^{-1}(\beta_2 + a_2u_2 + a_1 + a_2 + 1))}=1,$$ 
                which leads to
                    $$a_2(a_1+a_2+\overline{\beta_2}+1)u_2^2 + (\beta_2\overline{\beta_2}+(\beta_2+\overline{\beta_2})(a_1+a_2+1)+1)u_2 + a_2(a_1+a_2+\beta_2+1)=0.$$
                When $a_2\neq0$ and $a_1+a_2+\overline{\beta_2}+1\neq0$, this reduces to
                \begin{equation}\label{eqt:meq1_qeven_le3_2}
                    u_2^2 + \frac{(\beta_2\overline{\beta_2}+(\beta_2+\overline{\beta_2})(a_1+a_2+1)+1)}{a_2(a_1+a_2+\overline{\beta_2}+1)}u_2 + \frac{a_1+a_2+\beta_2+1}{a_1+a_2+\overline{\beta_2}+1}=0.
                \end{equation}
                A straightforward computation shows that
                    $$\tr^{2s}_1\left(\frac{\frac{a_1+a_2+\beta_2+1}{a_1+a_2+\overline{\beta_2}+1}}{\frac{(\beta_2\overline{\beta_2}+(\beta_2+\overline{\beta_2})(a_1+a_2+1)+1)^2}{a_2^2(a_1+a_2+\overline{\beta_2}+1)^2}}\right) = \tr^{2s}_1\left(\frac{a_2^2(a_1+a_2+\beta_2+1)(a_1+a_2+\overline{\beta_2}+1)}{(\beta_2\overline{\beta_2}+(\beta_2+\overline{\beta_2})(a_1+a_2+1)+1)^2}\right) = 0,$$
                and 
                    $$\frac{a_1+a_2+\beta_2+1}{a_1+a_2+\overline{\beta_2}+1}=\frac{\frac{(\beta_2\overline{\beta_2}+(\beta_2+\overline{\beta_2})(a_1+a_2+1)+1)}{a_2(a_1+a_2+\overline{\beta_2}+1)}}{\overline{\left(\frac{(\beta_2\overline{\beta_2}+(\beta_2+\overline{\beta_2})(a_1+a_2+1)+1)}{a_2(a_1+a_2+\overline{\beta_2}+1)}\right)}}.$$
                By Lemma \ref{lem:sol_in_unit_peq2}, \eqref{eqt:meq1_qeven_le3_2} possesses solutions in $U_{q+1}$ if and only if there exists $a_1,a_2$ such that
                    $$\tr^{s}_1\left(\frac{a_2^2(a_1+a_2+\beta_2+1)(a_1+a_2+\overline{\beta_2}+1)}{(\beta_2\overline{\beta_2}+(\beta_2+\overline{\beta_2})(a_1+a_2+1)+1)^2}\right) = 1.$$
                Note that
                    \begin{align*}
                        &\sum\limits_{\substack{a_1,a_2\in\gf_q^*\\a_1+a_2+\beta_2+1\neq0\\\beta_2\overline{\beta_2}+(\beta_2+\overline{\beta_2})(a_1+a_2+1)+1\neq0}}(-1)^{\tr^s_1\left(\frac{a_2^2(a_1+a_2+\beta_2+1)(a_1+a_2+\overline{\beta_2}+1)}{(\beta_2\overline{\beta_2}+(\beta_2+\overline{\beta_2})(a_1+a_2+1)+1)^2}\right)}\\
                        =&\sum\limits_{\substack{a_1,a_2\in\gf_q^*\\\beta_2\overline{\beta_2}+(\beta_2+\overline{\beta_2})(a_1+a_2+1)+1\neq0}}(-1)^{\tr^s_1\left(\frac{a_2^2(a_1+a_2+\beta_2+1)(a_1+a_2+\overline{\beta_2}+1)}{(\beta_2\overline{\beta_2}+(\beta_2+\overline{\beta_2})(a_1+a_2+1)+1)^2}\right)}-\sum\limits_{\substack{a_1,a_2\in\gf_q^*\\a_1+a_2+\beta_2+1=0\\\beta_2\overline{\beta_2}+(\beta_2+\overline{\beta_2})(a_1+a_2+1)+1\neq0}}1\\
                        =&\sum\limits_{\substack{a_1,a_2\in\gf_q\\\beta_2\overline{\beta_2}+(\beta_2+\overline{\beta_2})(a_1+a_2+1)+1\neq0}}(-1)^{\tr^s_1\left(\frac{a_2^2(a_1+a_2+\beta_2+1)(a_1+a_2+\overline{\beta_2}+1)}{(\beta_2\overline{\beta_2}+(\beta_2+\overline{\beta_2})(a_1+a_2+1)+1)^2}\right)}-\sum\limits_{\substack{a_1,a_2\in\gf_q^*\\a_1+a_2+\beta_2+1=0\\\beta_2\overline{\beta_2}+(\beta_2+\overline{\beta_2})(a_1+a_2+1)+1\neq0}}1\\
                        &-\sum\limits_{\substack{a_1\in\gf_q^*\\\beta_2\overline{\beta_2}+(\beta_2+\overline{\beta_2})(a_1+1)+1\neq0}}1
                         - \sum\limits_{\substack{a_2\in\gf_q^*\\\beta_2\overline{\beta_2}+(\beta_2+\overline{\beta_2})(a_2+1)+1\neq0}}(-1)^{\tr^s_1\left(\frac{a_2^2(a_2+\beta_2+1)(a_2+\overline{\beta_2}+1)}{(\beta_2\overline{\beta_2}+(\beta_2+\overline{\beta_2})(a_2+1)+1)^2}\right)}-1\\
                        =& -2q+3.
                    \end{align*}
                The last equality follows from the two calculations below, which are straightforward to verify:
                    $$\left\lvert\{a : a\in\gf_q^*,~\beta\overline{\beta}+(\beta+\overline{\beta})(a+1)+1\neq0\}\right\rvert=\begin{cases}
                        q-1, & \text{ when } \beta\in\gf_q\setminus\{1\}, \\
                        q-2, & \text{ when } \beta\not\in\gf_q,\\
                    \end{cases}$$
                and \begin{align*}
                        &\left\lvert\{(a_1,a_2) : a_1,a_2\in\gf_q^*,~a_1+a_2+\beta+1=0,\beta\overline{\beta}+(\beta+\overline{\beta})(a_1+a_2+1)+1\neq0\}\right\rvert\\
                        =&\begin{cases}
                            q-2, & \text{ when } \beta\in\gf_q\setminus\{1\},\\
                            0,   & \text{ otherwise}.
                        \end{cases}
                    \end{align*}
                Finally, it holds that
                    \begin{align*}
                        &\sum\limits_{\substack{a_1,a_2\in\gf_q^*\\a_1+a_2+\beta_2+1\neq0\\\beta_2\overline{\beta_2}+(\beta_2+\overline{\beta_2})(a_1+a_2+1)+1\neq0}}\frac{1-(-1)^{\tr^s_1\left(\frac{a_2^2(a_1+a_2+\beta_2+1)(a_1+a_2+\overline{\beta_2}+1)}{(\beta_2\overline{\beta_2}+(\beta_2+\overline{\beta_2})(a_1+a_2+1)+1)^2}\right)}}{2}\\
                        =&\frac{1}{2}\left[\sum\limits_{\substack{a_1,a_2\in\gf_q^*\\a_1+a_2+\beta_2+1\neq0\\\beta_2\overline{\beta_2}+(\beta_2+\overline{\beta_2})(a_1+a_2+1)+1\neq0}}1-\sum\limits_{\substack{a_1,a_2\in\gf_q^*\\a_1+a_2+\beta_2+1\neq0\\\beta_2\overline{\beta_2}+(\beta_2+\overline{\beta_2})(a_1+a_2+1)+1\neq0}}(-1)^{\tr^s_1\left(\frac{a_2^2(a_1+a_2+\beta_2+1)(a_1+a_2+\overline{\beta_2}+1)}{(\beta_2\overline{\beta_2}+(\beta_2+\overline{\beta_2})(a_1+a_2+1)+1)^2}\right)}\right]\\
                        =& \frac{q^2-q}{2}.
                    \end{align*}
                Thus, when $q>2$ is even, a solution to system \eqref{eqtsys:meq1_qeven_le3_4} is given by
                    $$\begin{cases}
                        a_1,a_2 \in\gf_q^* \text{ satisfying } a_1+a_2+\beta_2+1\neq0 \text{ and } \beta_2\overline{\beta_2}+(\beta_2+\overline{\beta_2})(a_1+a_2+1)+1\neq0 \\
                        \text{ such that } \tr^{s}_1\left(\frac{a_2^2(a_1+a_2+\beta_2+1)(a_1+a_2+\overline{\beta_2}+1)}{(\beta_2\overline{\beta_2}+(\beta_2+\overline{\beta_2})(a_1+a_2+1)+1)^2}\right) = 1, \\
                        a_3 = 1 + a_1 + a_2,\\
                        u_1=a_1^{-1}(\beta_2 + a_2u_2 + a_1 + a_2 + 1), \\
                        u_2\in U_{q+1} \text{ satisfying } \eqref{eqt:meq1_qeven_le3_2},\\
                        u_3 = 1.
                    \end{cases}
                    $$
    \end{itemize}
    Having treated both cases, we complete the proof.
\end{proof}

The covering radius of $\left(\C_{(q,q+1,3,0)}\right)$ for even $q$ is given in the following theorem.
\begin{theorem}\label{thm:meq1_qeven}
    When $q$ is even and $q>2$, it holds that $\rho\left(\C_{(q,q+1,3,0)}\right)=3$.
\end{theorem}
\begin{proof}
    Based on Lemma \ref{lem:CR_PCmatrix}, we obtain $\rho\bigl(\mathcal{C}_{(q,q+1,3,0)}\bigr) \geqslant 3$ from Lemma \ref{lem:meq1_qeven_lbound}, while Lemma \ref{lem:meq1_qeven_ubound} implies $\rho\bigl(\mathcal{C}_{(q,q+1,3,0)}\bigr) \leqslant 3$. Hence, the equality holds.
\end{proof}

\section{The covering radius of $\C_{(q,q^m+1,3,0)}$ with $m>1$}\label{sec:CR_0}
This section is devoted to determining the covering radii of $\C_{(q,q^m+1,3,0)}$ for $m>1$. We treat the cases of odd and even \(q\) separately. It is known from \cite{CaoCXW2025BCHq_2} that the code \(C_{(q,q^m+1,3,0)}\) has parameters $[q^m+1,q^m-2m,d]$, where $d\geqslant4$ and $d=4$ when $q$ is odd. Moreover, its extension-field parity-check matrix can be written as
        $$H_{(q,q^m+1,3,0)}=\begin{pmatrix}
                1 & 1 & 1 & \cdots & 1 \\
                1 & \mu^{1} & \mu^{2} & \cdots & \mu^{q^m} 
            \end{pmatrix},$$
where \(\mu\) is a primitive \((q^m+1)\)-th root of unity and \(U_{q^m+1}=\langle \mu\rangle\).
Our analysis starts with a lemma establishing a lower bound. 

\begin{lemma}\label{lem:any_ge3}
    Let $s$ be a positive integer, $p$ be a prime and $q=p^s$. When $p=2$ and $m>1$ or $p>2$ and $m>1$ is even, there exist $\beta_1 \in\gf_q,\beta_2 \in\gf_{q^{2m}}$ such that the following system has no solution $(a_1,a_2,u_1,u_2)\in(\gf_q)^2\times(U_{q^m+1})^2$:
    \begin{equation}\label{eqtsys:ge3}
        \begin{cases}
            \begin{aligned}
                &a_1 + a_2 &=& \beta_1, \\
                &a_1u_1 + a_2u_2 &=& \beta_2.
            \end{aligned}
        \end{cases}
    \end{equation}
\end{lemma}
\begin{proof}
    The proof distinguishes between even and odd characteristic.
    \begin{itemize}
        \item If $p=2$, then we tend to prove this lemma by setting $\beta_1 \neq 0$ and $\beta_2 = 0$. First, note that a solution with $a_2 = 0$ would force $a_1 = 0$ (from $\beta_2=0$) and then $\beta_1 = 0$, which is impossible. Therefore, $a_2 \neq 0$ in any admissible solution. From the second equation in system (\ref{eqtsys:ge3}) we have $u_2=a_2^{-1}a_1u_1$. The condition $u_2 \in U_{q^m+1}$ is equivalent to $(a_2^{-1}a_1)^2 = 1$, whence $a_1 = a_2$. However, with $a_1 = a_2$, the first equation gives $\beta_1 = 0$, a contradiction.
        \item If $p$ is odd, denote $(\gf_q^*)^2$ the square elements in $\gf_q^*$. The proof proceeds by setting $\beta_1 = 0$ and $\overline{\beta_2}\beta_2$ be a non-square in $\gf_q^*$. Note that such $\beta_2$ indeed exists since the norm map $N_{\mathbb F_{q^{2m}}/\mathbb F_{q^m}}:\mathbb F_{q^{2m}}^*\longrightarrow \mathbb F_{q^m}^*,~x\longmapsto x^{q^m+1},$ is surjective.
        When $\beta_1 = 0$, the first equation in \eqref{eqtsys:ge3} leads to $a_2=-a_1$.Substituting this into the second equation gives
        \begin{equation}\label{eqt:ge3_1}
            a_1(u_1-u_2)=\beta_2.
        \end{equation}
        From \eqref{eqt:ge3_1} we obtain $u_1=u_2+a_1^{-1}\beta_2$.
        The condition $u_1\in U_{q^m+1}$ then requires 
            $$(u_2+a_1^{-1}\beta_2)(u_2^{-1}+a_1^{-1}\overline{\beta_2})=1,$$
        which can be simplified to
            \begin{equation}\label{eqt:ge3_2}
                u_2^2+a_1^{-1}\beta_2u_2+\frac{\beta_2}{\overline{\beta_2}}=0.
            \end{equation}
        Note that the discriminant of \eqref{eqt:ge3_2}, namely $\beta_2^{1-q^m}(a_1^{-2}\beta_2 \overline{\beta_2} - 4)$, is a square in $\mathbb{F}_{q^{2m}}$. By Lemma \ref{lem:sol_in_unit_p}, equation \eqref{eqt:ge3_2} admits solutions in $U_{q^m+1}$ only if
            $$\chi_{q^m}\left( \frac{\beta_2\overline{\beta_2} - 4a_1^2}{\beta_2\overline{\beta_2}} \right)=-1 \text{ or } \beta_2\overline{\beta_2} - 4a_1^2=0.$$
        However, since $m$ is even, the defining condition of $\overline{\beta_2}\beta_2\in\gf_q^*\setminus (\gf_q^*)^2$ implies $\chi_{q^m}\left( \frac{\beta_2\overline{\beta_2} - a}{\beta_2\overline{\beta_2}} \right) = 1 $ for all $a \in (\gf_q^*)^2$, which contradicts the necessary condition above. Therefore, \eqref{eqtsys:ge3} admits no solutions under given conditions.
    \end{itemize}
\end{proof}

Lemma \ref{lem:any_ge3} and Lemma \ref{lem:CR_PCmatrix} yield a lower bound directly.
\begin{corollary}\label{col:any_ge3}
    We obtain the lower bound $\rho\left(\C_{(q,q^m+1,3,0)}\right)\geqslant3$ for $p=2,m>1$ and $p>2$, $m$ is even. 
\end{corollary}

We then proceed to derive the exact covering radii, treating the cases of even $p$ and odd $p$ separately.

\subsection{The covering radius of $\C_{(q,q^m+1,3,0)}$ for even $q$ and $m>1$}\label{subsec:CR_even_mge1}
In this subsection, we investigate the covering radii of $\C_{(q,q^m+1,3,0)}$ for even $q$ and $m>1$. We first treat the case \(q=2\). Lemma \ref{lem:e_seq1_lbound} provides the required lower bounds, while Lemmas \ref{lem:e_seq1_meven_ubound} and \ref{lem:e_seq1_modd_ubound} establish the corresponding upper bounds for even and odd \(m\), respectively. These results lead to Theorems \ref{thm:e_seq1_meven} and \ref{thm:e_seq1_modd}.

The lemma below states the lower bound on covering radii of $\C_{(q,q^m+1,3,0)}$ when $s=1$ and $m>1$.
\begin{lemma}\label{lem:e_seq1_lbound}
    When $p=2$, $s=1$(i.e., $q=2$) and $m>1$, the following statements hold:
    \begin{itemize}
        \item When $m$ is even, there exist $\beta_1\in\gf_q,\beta_2\in\gf_{q^{2m}}$ such that the following system has no solution $(a_1,a_2,a_3,a_4,u_1,u_2,u_3,u_4)\in(\gf_q)^4\times(U_{q^m+1})^4$:
                \begin{equation}\label{eqtsys:e_1_ge5}
                    \begin{cases}
                        \begin{aligned}
                            &a_1 + a_2 + a_3 + a_4 &=& \beta_1, \\
                            &a_1u_1 + a_2u_2 + a_3u_3 + a_4u_4  &=& \beta_2.
                        \end{aligned}
                    \end{cases}
                \end{equation}
        \item When $m$ is odd, there exist $\beta_1\in\gf_q,\beta_2 \in\gf_{q^{2m}}$ such that the following system has no solution $(a_1,a_2,a_3,u_1,u_2,u_3)\in(\gf_q)^3\times(U_{q^m+1})^3$:
                \begin{equation}\label{eqtsys:e_1_ge4}
                    \begin{cases}
                        \begin{aligned}
                            &a_1 + a_2 + a_3 &=& \beta_1, \\
                            &a_1u_1 + a_2u_2 + a_3u_3  &=& \beta_2.
                        \end{aligned}
                    \end{cases}
                \end{equation}
    \end{itemize}
\end{lemma}
\begin{proof}
    \begin{itemize} 
        \item To prove the first statement, where $m$ is even, take $\beta_1=1$ and $\beta_2=0$. In this case, the number of $1$'s in the tuple $(a_1,a_2,a_3,a_4)$ must be odd. If there is exactly one $1$, we may assume without loss of generality that $a_1=1$ and $a_2=a_3=a_4=0$. The second equation in (\ref{eqtsys:e_1_ge5}) then gives $u_1=0$, contradicting to $u_1\in U_{q^m+1}$. If there are exactly three $1$'s, assume W.L.O.G. that $a_1=a_2=a_3=1$ and $a_4=0$, then the second equation in (\ref{eqtsys:e_1_ge5}) yields
            $$u_1+u_2+u_3=0.$$
        Due to $u_1\in U_{q^m+1}$, $u_1=u_2+u_3$ is a solution if and only if 
        \begin{equation}\label{eqt:e_1_ge5_1}
            (u_2+u_3)(u_2^{-1}+u_3^{-1})=1,
        \end{equation}
        (\ref{eqt:e_1_ge5_1}) can be simplified to
        \begin{equation}\label{eqt:e_1_ge5_1_2}
            u_2^2+u_3u_2+u_3^2=0.
        \end{equation}
        Then it suffices to show that for any $u_3\in U_{q^m+1}$ (\ref{eqt:e_1_ge5_1_2}) admits no solution in $U_{q^m+1}$. According to Lemma \ref{lem:sol_in_unit_peq2}, (\ref{eqt:e_1_ge5_1_2}) has solutions in $U_{q^m+1}$ requires $\tr^m_1(1)=1$, which is impossible since $m$ is even.
        In short, the system (\ref{eqtsys:e_1_ge5}) has no solution when $\beta_1=1$ and $\beta_2=0$.
        \item For the statement on $m$ is odd, we prove this statement by setting $\beta_1=0$ and $\beta_2\in\gf_{q^m}^*$ satisfying $\tr_1^m(\beta_2^{-1})=0$. Since $\beta_1=0$ and $\beta_2\neq0$, W.L.O.G. we may assume that $a_1=a_2=1$ and $a_3=0$. Then the second equation in (\ref{eqtsys:e_1_ge5}) then gives
            $$u_1+u_2=\beta_2.$$
        Note that $u_2=u_1+\beta_2$ lies in $U_{q^m+1}$ if and only if 
            $$(u_1+\beta_2)(u_1^{-1}+\beta_2)=1,$$
        which can be rewritten as
        \begin{equation}\label{eqt:e_1_ge4_2}
            \beta_2u_1^2+\beta_2^2u_1+\beta_2=0.
        \end{equation}
        Since $\beta_2\neq0$, (\ref{eqt:e_1_ge4_2}) can be reduced to
        \begin{equation}\label{eqt:e_1_ge4_3}
            u_1^2+\beta_2u_1+1=0.
        \end{equation}
        According to Lemma \ref{lem:sol_in_unit_peq2}, (\ref{eqt:e_1_ge4_3}) has solutions in $U_{q^m+1}$ requires $\tr^m_1\left(\frac{1}{\beta_2^2}\right)=\tr^m_1\left(\frac{1}{\beta_2}\right)=1$, a contradiction to the condition $\tr_1^m(\beta_2^{-1})=0$.
    \end{itemize}
\end{proof}

The lemma below states the upper bound on covering radius of $\C_{(q,q^m+1,3,0)}$ when $s=1$ and $m>1$ is even. For brevity, the proof is relegated to the appendix.
\begin{lemma}\label{lem:e_seq1_meven_ubound}
    Let $s=1$, $p=2$(i.e. $q=2$) and $m\geqslant4$ be an even positive integer. For every $\beta_1\in\gf_q$, $\beta_2\in\gf_{q^{m}}$, there exist $a_i\in\gf_q$ and $u_i\in U_{q^m+1}$ with $i=1,2,3,4,5$ such that the following equation holds:
       \begin{equation}\label{eqtsys:e_seq1_meven}
            \begin{cases}
                \begin{aligned}
                    &a_1 + a_2 + a_3 + a_4 + a_5 &=& \beta_1, \\
                    &a_1u_1 + a_2u_2 + a_3u_3 + a_4u_4 + a_5u_5 &=& \beta_2.
                \end{aligned}
            \end{cases}
        \end{equation}
\end{lemma}

The following theorem establishes the value of the covering radius of $\C_{(q,q^m+1,3,0)}$, in the case where $s=1$ and $m>1$ is even.
\begin{theorem}\label{thm:e_seq1_meven}
    When $p=2$, $s=1$(i.e. $q=2$) and $m>1$ is even, $\rho\left(\C_{(q,q^m+1,3,0)}\right)=5$.
\end{theorem}
\begin{proof}
    Recall that the parameters of $\C_{(q,q^m+1,3,0)}$ is $[q^m+1,q^m-2m,d]$ with $d\geqslant3$. To prove $\rho\left(\C_{(q,q^m+1,3,0)}\right)\leqslant5$, according to Lemma \ref{lem:CR_PCmatrix}, we just need to prove that for every $\beta_1\in\gf_q,\beta_2\in\gf_{q^{2m}}$, there exist $a_i\in\gf_q$ and $u_i\in U_{q^m+1}$($i=1,2,3,4,5$) such that the following system holds:
        \begin{equation}\label{eqtsys:e_seq1_meven_0}
            \begin{cases}
                \begin{aligned}
                    &a_1 + a_2 + a_3 + a_4 + a_5&=& \beta_1, \\
                    &a_1u_1 + a_2u_2 + a_3u_3 + a_4u_4 + a_5u_5&=& \beta_2.
                \end{aligned}
            \end{cases}
        \end{equation}
    Given the representation of every $\beta_2 = u\beta$ where $u \in U_{q^m+1}$ and $\beta \in \gf_{q^m}$ when $p=2$, it suffices to prove that for every $\beta_1\in\gf_q$, $\beta \in \gf_{q^m}$, the following system admits a solution $(a_1,a_2,a_3,a_4,a_5,$ $u_1,u_2,u_3,u_4,u_5)$ in $(\gf_q)^5\times U_{q^m+1}^5$.
        \begin{equation}
            \begin{cases}
                \begin{aligned}\label{eqtsys:e_seq1_meven_01}
                    &a_1 + a_2 + a_3 + a_4 + a_5&=& \beta_1, \\
                    &a_1u_1 + a_2u_2 + a_3u_3 + a_4u_4 + a_5u_5&=& \beta.
                \end{aligned}
            \end{cases}
        \end{equation}
    By Lemma \ref{lem:e_seq1_meven_ubound}, equation \eqref{eqtsys:e_seq1_meven_01} is solvable for \( m \geqslant 4 \); hence, the upper bound \(\rho(\C_{(q,q^m+1,3,0)}) \leqslant 5\) holds in this case. Together with the matching lower bound from Lemma \ref{lem:e_seq1_lbound}, this proves the equality for these parameters. For the exceptional case \(m=2\), a direct computation in MAGMA using the intrinsic function \texttt{CoveringRadius} confirms that \(\rho(\C_{(2,5,3,0)}) = 5\). In summary, for all \(q = 2\) and integers \(m > 1\) being even, we have \(\rho(\C_{(q,q^m+1,3,0)}) = 5\).
\end{proof}

The lemma below states the upper bound on covering radius of $\C_{(q,q^m+1,3,0)}$ when $s=1$ and $m>1$ is odd. Since the proof is analogous to that of Lemma \ref{lem:e_seq1_meven_ubound}, we omit it.
\begin{lemma}\label{lem:e_seq1_modd_ubound}
    Let $s=1$, $p=2$(i.e. $q=2$) and $m\geqslant5$ be an odd positive integer. For every $\beta_1\in\gf_q$, $\beta_2\in\gf_{q^m}$, there exist $a_i\in\gf_q$ and $u_i\in U_{q^m+1}$ with $i=1,2,3,4$ such that the following equation holds:
       \begin{equation}\label{eqtsys:e_seq1_modd}
            \begin{cases}
                \begin{aligned}
                    &a_1 + a_2 + a_3 + a_4 &=& \beta_1, \\
                    &a_1u_1 + a_2u_2 + a_3u_3 + a_4u_4 &=& \beta_2.
                \end{aligned}
            \end{cases}
        \end{equation}
\end{lemma}

The following theorem establishes the value of the covering radius of $\C_{(q,q^m+1,3,0)}$, in the case where $s=1$ and $m>1$ is odd.
\begin{theorem}\label{thm:e_seq1_modd}
    When $p=2$, $s=1$(i.e. $q=2$) and $m>1$ is odd, $\rho\left(\C_{(q,q^m+1,3,0)}\right)=4$.
\end{theorem}
\begin{proof}
    Recall that the parameters of $\C_{(q,q^m+1,3,0)}$ is $[q^m+1,q^m-2m,d]$ with $d\geqslant3$. To prove \(\rho\left(\C_{(q,q^m+1,3,0)}\right)\leqslant4\), according to Lemma \ref{lem:CR_PCmatrix}, we just need to prove that for every $\beta_1\in\gf_q,\beta_2\in\gf_q^{2m}$, there exist $a_1,a_2,a_3,a_4\in\gf_q$ and $u_1,u_2,u_3,u_4\in U_{q^m+1}$ such that the following system holds:
        \begin{equation}
            \begin{cases}
                \begin{aligned}\label{eqtsys:e_seq1_modd_0}
                    &a_1 + a_2 + a_3 + a_4&=& \beta_1, \\
                    &a_1u_1 + a_2u_2 + a_3u_3 + a_4u_4&=& \beta_2.
                \end{aligned}
            \end{cases}
        \end{equation}
    Given the representation of every $\beta_2 = u\beta$ where $u \in U_{q^m+1}$ and $\beta \in \gf_{q^m}$ when $p=2$, it suffices to prove that for every $\beta_1\in\gf_q$ and every $\beta \in \gf_{q^m}$, the following system admits a solution $(a_1,a_2,a_3,a_4,u_1,u_2,$ $u_3,u_4)$ in $(\gf_q)^4\times (U_{q^m+1})^4$.
        \begin{equation}
            \begin{cases}
                \begin{aligned}\label{eqtsys:e_seq1_modd_01}
                    &a_1 + a_2 + a_3 + a_4&=& \beta_1, \\
                    &a_1u_1 + a_2u_2 + a_3u_3 + a_4u_4&=& \beta.
                \end{aligned}
            \end{cases}
        \end{equation}
    By Lemma \ref{lem:e_seq1_modd_ubound}, equation \eqref{eqtsys:e_seq1_modd_01} is solvable for \( m \geqslant 5 \); hence, the upper bound \(\rho(\C_{(q,q^m+1,3,0)}) \leqslant 4\) holds in this case. Together with the matching lower bound from Lemma \ref{lem:e_seq1_lbound}, this proves the equality for these parameters. For the exceptional case \(m=3\), a direct computation in MAGMA using the intrinsic function \texttt{CoveringRadius} confirms that \(\rho(\C_{(2,9,3,0)}) = 4\). In summary, for all \(q = 2\) integers \(m > 1\) being odd, we have \(\rho(\C_{(q,q^m+1,3,0)}) = 4\).
\end{proof}

Next, we turn to the case \( s > 1 \). Analogous to the \( s = 1 \) case, the analysis begins with the equations in Lemma \ref{lem:e_sge2_modd_ubound}, \ref{lem:e_sge2_meven_lbound} and \ref{lem:e_sge2_meven_ubound}; studying their solutions yields the covering radii for even \( q \), \( s > 1 \), and \( m > 1 \), as stated in Theorem \ref{thm:e_sge2_modd} and \ref{thm:e_sge2_meven}.

The lemma below states the upper bound on covering radius of $\C_{(q,q^m+1,3,0)}$ when $s>1$ and $m>1$ is odd. For brevity, the proof is relegated to the appendix.
\begin{lemma}\label{lem:e_sge2_modd_ubound}
    Let $p=2$, $s>1$ and $m>1$ be an odd positive integer. Then the following two equations are solvable.
    \begin{itemize}
        \item For every $\beta\in\gf_{q^{m}}^*$, set $b \in \gf_q\setminus\{0, 1\}$ and if $1+\beta\in\gf_q$ additionally $b\neq1+\beta$, then there exist $u_i\in U_{q^m+1}$ with $i=1,2,3$ such that the following equation holds:
            \begin{equation}\label{eqtsys:e_sge1_modd_0}
                bu_1 + (b+1)u_2 + u_3  = \beta.
            \end{equation}
        \item For every $\beta\in\gf_{q^{m}}$, set $a\in\gf_{q}$ with $\tr^{sm}_1(a)=1$ then there exist $u_i\in U_{q^m+1}$ with $i=1,2,3$ such that the following equation holds:
            \begin{equation}\label{eqtsys:e_sge1_modd_00}
                au_1 + au_2 + u_3  = \beta.
            \end{equation}
    \end{itemize}
\end{lemma}

The following theorem establishes the value of the covering radius of $\C_{(q,q^m+1,3,0)}$, in the case where $s>1$ and $m>1$ is odd.
\begin{theorem}\label{thm:e_sge2_modd}
    When $p=2$, $s>1$ and $m>1$ is odd, $\rho\left(\C_{(q,q^m+1,3,0)}\right)=3$.
\end{theorem}
\begin{proof}
    Recall that the parameters of $\C_{(q,q^m+1,3,0)}$ is $[q^m+1,q^m-2m,d]$ with $d\geqslant3$. To prove $\rho\left(\C_{(q,q^m+1,3,0)}\right)\leqslant3$ , according to Lemma \ref{lem:CR_PCmatrix}, we just need to prove that for every $\beta_1\in\gf_q,\beta_2\in\gf_q^{2m}$, there exist $a_1,a_2,a_3\in\gf_q$ and $u_1,u_2,u_3\in U_{q^m+1}$ such that the following system holds:
        \begin{equation}\label{eqtsys:even_sge1}
            \begin{cases}
                \begin{aligned}
                    &a_1 + a_2 + a_3 &=& \beta_1, \\
                    &a_1u_1 + a_2u_2 + a_3u_3 &=& \beta_2.
                \end{aligned}
            \end{cases}
        \end{equation}
    The proof of this lemma proceeds by considering two cases: $\beta_1=0$ and $\beta_1\neq0$.
    \begin{itemize}
        \item For $\beta_1=0$, if $\beta_2=0$ then the solutions are $a_1=a_2=a_3=0$ with $u_1,u_2,u_3 \in U_{q^m+1}$. Otherwise ($\beta_2 \neq 0$), set $b \in \gf_q\setminus\{0, 1, 1+\beta\}$ then the choice $a_1=b$, $a_2=b+1$ and $a_3=1$ transforms (\ref{eqtsys:even_sge1}) into
            \begin{equation}\label{eqtsys:even_sge1_0}
                bu_1 + (1+b)u_2 + u_3 = \beta_2.
            \end{equation}
            Given the unique representation $\beta_2 = u\beta$ where $u \in U_{q^m+1}$ and $\beta \in \gf_{q^m}^*$, it suffices to prove that for every $\beta \in \gf_{q^m}^*$, the following equation admits a solution $(u_1,u_2,u_3)$ in $U_{q^m+1}^3$.
                \begin{equation}\label{eqtsys:even_sge1_1}
                    bu_1 + (1+b)u_2 + u_3 = \beta.
                \end{equation}
            By Lemma \ref{lem:e_sge2_modd_ubound}, solvability of (\ref{eqtsys:even_sge1_0}) holds under the conditions $s>1$ and odd $m>1$.
        \item When $\beta_1\neq0$, given the unique representation $\beta_2 = u\beta$ where $u \in U_{q^m+1}$ and $\beta \in \gf_{q^m}$, (\ref{eqtsys:even_sge1}) can be reduced to
            \begin{equation}\label{eqtsys:even_sge1_5}
                \begin{cases}
                    \begin{aligned}
                        &\frac{a_1}{\beta_1} + \frac{a_2}{\beta_1} + \frac{a_3}{\beta_1} &=& 1, \\
                        &\frac{a_1}{\beta_1}\frac{u_1}{u} + \frac{a_2}{\beta_1}\frac{u_2}{u} + \frac{a_3}{\beta_1}\frac{u_3}{u} &=& \frac{\beta_2}{\beta_1u}.
                    \end{aligned}
                \end{cases}
            \end{equation}
            It therefore remains to prove that for every $\beta\in\gf_{q^{m}}$, there exist $b_1,b_2,b_3\in\gf_q$ and $u_1,u_2,u_3\in U_{q^m+1}$ such that the following system holds:
                \begin{equation}\label{eqtsys:even_sge1_6}
                    \begin{cases}
                        \begin{aligned}
                            &b_1 + b_2 + b_3 &=& 1, \\
                            &b_1u_1 + b_2u_2 + b_3u_3 &=& \beta.
                        \end{aligned}
                    \end{cases}
                \end{equation}
                Upon setting $b_1=b_2=a$ and $b_3=1$ where $a\in\gf_{q}$ with $\tr^{sm}_1(a)=1$, system (\ref{eqtsys:even_sge1_6}) reduces to
                    \begin{equation}\label{eqtsys:even_sge1_7}
                        au_1 + au_2 + u_3 = \beta.
                    \end{equation}
                By Lemma \ref{lem:e_sge2_modd_ubound}, solvability of (\ref{eqtsys:even_sge1_6}) holds under the conditions $s>1$ and odd $m>1$.
    \end{itemize}
    Thus, the upper bound \(\rho(\C_{(q,q^m+1,3,0)}) \leqslant 3\) holds in this case. Together with the matching lower bound from Corollary \ref{col:any_ge3}, this proves \(\rho(\C_{(q,q^m+1,3,0)}) = 3\) when $s>1$ and $m>1$ is odd.
\end{proof}

The lemma below states the lower bound on covering radius of $\C_{(q,q^m+1,3,0)}$ when $s>1$ and $m>1$ is even.
\begin{lemma}\label{lem:e_sge2_meven_lbound}
    Let $p=2$, $s>1$ be a positive integer and $q=p^s$. When $m\in\Z^{+}$ is even, there exist $\beta_1 \in\gf_q,\beta_2 \in\gf_{q^{2m}}$ such that for all $a_i\in\gf_q$ and $u_i\in U_{q^m+1}$ for $i=1,2,3$, the following system has no solution.
        \begin{equation}\label{eqtsys:e_sge2_ge4}
            \begin{cases}
                \begin{aligned}
                    &a_1 + a_2 + a_3 &=& \beta_1, \\
                    &a_1u_1 + a_2u_2 + a_3u_3 &=& \beta_2.
                \end{aligned}
            \end{cases}
        \end{equation}
\end{lemma}
\begin{proof}
    We tend to prove this lemma by setting $\beta_1=1$ and $\beta_2=0$.
    Solving the first equation in system (\ref{eqtsys:e_sge2_ge4}) for $a_1$ gives $a_1 = a_2 + a_3 + 1$ which is obvious in $\gf_q$. Substitution of this into the second equation yields
    \begin{equation}\label{eqt:e_sge2_ge4_1}
        (a_2+a_3+1)u_1+a_2u_2+a_3u_3=0.
    \end{equation}
    \begin{itemize}
        \item If the system (\ref{eqtsys:e_sge2_ge4}) admits a solution of the form $a_2+a_3+1=0$, then it follows from (\ref{eqt:e_sge2_ge4_1}) that
                \begin{equation}\label{eqt:e_sge2_ge4_2}
                    a_2u_2+a_3u_3=0.
                \end{equation}
                If $a_2=0$ is a solution, then (\ref{eqt:e_sge2_ge4_2}) would reduce to $a_3u_3=0$, implying $a_3=0$. Consequently, we would have $1+a_2+a_3=1$, which contradicts to the condition $a_2+a_3+1=0$. Therefore, $a_2\neq0$. From (\ref{eqt:e_sge2_ge4_2}) we then obtain $u_2=a_2^{-1}a_3u_3$. For such a $u_2\in U_{q^m+1}$ if and only if 
                \begin{equation}\label{eqt:e_sge2_ge4_3}
                    (a_2^{-1}a_3u_3)(a_2^{-1}a_3u_3^{-1})=1.
                \end{equation}
                Simplifying (\ref{eqt:e_sge2_ge4_3}) gives $a_2^2=a_3^2$; that is, $a_2=a_3$. This contradicts the condition $a_2+a_3+1=0$.
        \item If the system (\ref{eqtsys:e_sge2_ge4}) admits a solution of the form $a_2+a_3+1\neq0$, then it follows from (\ref{eqt:e_sge2_ge4_1}) that
                \begin{equation}\label{eqt:e_sge2_ge4_4}
                    u_1=\frac{a_2u_2+a_3u_3}{a_2+a_3+1}.
                \end{equation}
                The $u_1$ given by (\ref{eqt:e_sge2_ge4_4}) is in $U_{q^m+1}$ if and only if $\frac{a_2u_2+a_3u_3}{a_2+a_3+1}\frac{a_2u_2^{-1}+a_3u_3^{-1}}{a_2+a_3+1}=1$, which is equivalent to
                    $$a_2a_3(\frac{u_2}{u_3}+\frac{u_3}{u_2})+1=0.$$
                Set $u=\frac{u_2}{u_3}$. Then, as $u_2$ and $u_3$ range independently over $U_{q^m+1}$, so does $u$. Consequently, it suffices to show that the equation below admits no solution in $U_{q^m+1}$:
                \begin{equation}\label{eqt:e_sge2_ge4_5}
                    a_2a_3(u+u^{-1})+1=0.
                \end{equation}
                Obviously \eqref{eqt:e_sge2_ge4_5} holds only when $a_2a_3\neq0$ and $u\neq1$, then \eqref{eqt:e_sge2_ge4_5} gives
                $$\frac{1}{u+u^{-1}}=a_2a_3.$$
                Since $m$ is even, $\tr^{sm}_1(a_2a_3)=0$, this contradicts to Lemma \ref{lem:values_for_addsum}.
                Therefore, \eqref{eqt:e_sge2_ge4_5} is invalid.
    \end{itemize}
    To summarize, equation (\ref{eqt:e_sge2_ge4_1}) admits no solution any $a_2,a_3\in\gf_q$ and $u_i\in U_{q^m+1}(i=1,2,3)$. Thus, we conclude that the system (\ref{eqtsys:e_sge2_ge4}) is unsolvable when $\beta_1=1$ and $\beta_2=0$.
\end{proof}

The lemma below states the upper bound on covering radius of $\C_{(q,q^m+1,3,0)}$ when $s>1$ and $m>1$ is even. Since its proof is similar to that of Lemma \ref{lem:e_sge2_modd_ubound}, we omit it.
\begin{lemma}\label{lem:e_sge2_meven_ubound}
    Let $p=2$, $s>1$ be a positive integer and $q=p^s$. Let $\alpha_q$ be the primitive element of $\gf_q$. When $m\in\Z^{+}$ is even and $sm\geqslant6$, there exist $u_i\in U_{q^m+1}$ with $i=1,2,3,4$ such that the following equation holds:
        \begin{equation}\label{eqtsys:e_sge2_meven_le4}
           u_1 + u_2 + \alpha_q u_3 + (\alpha_q+1)u_4 = 0.
        \end{equation}
\end{lemma}

The following theorem establishes the value of the covering radius of $\C_{(q,q^m+1,3,0)}$, in the case where $s>1$ and $m>1$ is even.
\begin{theorem}\label{thm:e_sge2_meven}
    When $p=2$, $s>1$ and $m>1$ is even, $\rho\left(\C_{(q,q^m+1,3,0)}\right)=4$.
\end{theorem}
\begin{proof}
    Recall that the parameters of $\C_{(q,q^m+1,3,0)}$ is $[q^m+1,q^m-2m,d]$ with $d\geqslant3$. To prove $\rho\left(\C_{(q,q^m+1,3,0)}\right)\leqslant4$ , according to Lemma \ref{lem:CR_PCmatrix}, we just need to prove that for every $\beta_1\in\gf_q,\beta_2\in\gf_q^{2m}$, there exist $a_1,a_2,a_3,a_4\in\gf_q$ and $u_1,u_2,u_3,u_4\in U_{q^m+1}$ such that the following system holds:
        \begin{equation}\label{eqtsys:e_sge2_meven}
            \begin{cases}
                \begin{aligned}
                    &a_1 + a_2 + a_3 + a_4&=& \beta_1, \\
                    &a_1u_1 + a_2u_2 + a_3u_3 + a_4u_4 &=& \beta_2.
                \end{aligned}
            \end{cases}
        \end{equation}
    The proof of this theorem proceeds by considering two cases: $\beta_1=0$ and $\beta_1\neq0$.
    \begin{itemize}
        \item For $\beta_1=0$, if $\beta_2=0$ then the solutions are $a_1=a_2=a_3=a_4=0$ with $u_1,u_2,u_3 \in U_{q^m+1}$. Otherwise ($\beta_2 \neq 0$), the choice $a_1=a_2=a_3=a_4=1$ transforms (\ref{eqtsys:e_sge2_meven}) into
            \begin{equation}\label{eqt:e_sge2_meven_1}
                u_1 + u_2 + u_3 + u_4 = \beta_2.
            \end{equation}
            Given the unique representation $\beta_2 = u\beta$ where $u \in U_{q^m+1}$ and $\beta \in \gf_{q^m}^*$, our goal is to prove that for every $\beta \in \gf_{q^m}^*$, the following equation admits a solution $(u_1,u_2,u_3,u_4)$ in $U_{q^m+1}^4$.
                \begin{equation}\label{eqt:e_sge2_meven_2}
                    u_1 + u_2 + u_3 + u_4 = \beta.
                \end{equation}
            By the same analysis as case $\beta_1=0$ in Lemma \ref{lem:e_seq1_meven_ubound}, equation (\ref{eqt:e_sge2_meven_1}) is solvable when $sm\geqslant6$.
        \item When $\beta_1\neq0$, by the similar analysis to the case $m$ is odd when $p=2$, $s>1$, we just need to prove that for every $\beta\in\gf_{q^{m}}$, there exist $a_1,a_2,a_3,a_4\in\gf_q$ and $u_1,u_2,u_3\in U_{q^m+1}$ such that the following system holds:
                \begin{equation}\label{eqtsys:e_sge2_meven_2}
                    \begin{cases}
                        \begin{aligned}
                            &a_1 + a_2 + a_3 + a_4&=& 1, \\
                            &a_1u_1 + a_2u_2 + a_3u_3 + a_4u_4 &=& \beta.
                        \end{aligned}
                    \end{cases}
                \end{equation}
            \begin{itemize}
                \item  When $\beta\neq0$, the choice $a_1=a_2=a_3=1$ and $a_4=0$ transforms (\ref{eqtsys:e_sge2_meven_2}) into
                        \begin{equation}\label{eqt:e_sge2_meven_3}
                            u_1 + u_2 + u_3 = \beta.
                        \end{equation}
                    By the same analysis as the case $\beta_1=1$ in Lemma \ref{lem:e_seq1_meven_ubound}, equation (\ref{eqt:e_sge2_meven_3}) is solvable when $m\geqslant2$ and $\beta\neq0$.
                \item When $\beta=0$, the choice $a_1=a_2=1$, $a_3=\alpha_q$ and $a_4=\alpha_q+1$, where $\alpha_q$ is the generator element of $\gf_{q}$, transforms (\ref{eqtsys:e_sge2_meven_2}) into
                        \begin{equation}\label{eqtsys:even_sge1_4}
                            u_1 + u_2 + \alpha_q u_3 + (\alpha_q+1)u_4 = 0.
                        \end{equation}
                    By Lemma \ref{lem:e_sge2_meven_ubound}, solvability of (\ref{eqtsys:even_sge1_4}) holds under the conditions $sm\geqslant6$.
            \end{itemize}
    \end{itemize}
    Hence, the upper bound \(\rho(\C_{(q,q^m+1,3,0)}) \leqslant 4\) holds when $sm\geqslant6$. Together with the matching lower bound from Lemma \ref{lem:e_sge2_meven_lbound}, this proves the equality for these parameters. For the exceptional case \(s=2, m=2\), a direct computation in MAGMA using the intrinsic function \texttt{CoveringRadius} confirms that \(\rho(\C_{(4,17,3,0)}) = 4\). In summary, for all \(p = 2\), $s>1$ and integers \(m > 1\) being even, we have \(\rho(\C_{(q,q^m+1,3,0)}) = 4\).
\end{proof}

\subsection{The covering radius of $\C_{(q,q^m+1,3,0)}$ for odd $q$ and $m>1$}\label{subsec:CR_odd_mge1}
In this subsection, we turn to the case of odd $q$ and $m>1$ for the code $\C_{(q,q^m+1,3,0)}$. Similar to the even $q$ case, Lemmas \ref{lem:o_le3_1} and \ref{lem:o_le3_0} provide the necessary equations. These results collectively yield Theorems \ref{thm:podd_meven} and \ref{thm:podd_modd}, where the covering radii are investigated for odd $q$, $s \geqslant 1$, and $m>1$.

\begin{lemma}\label{lem:o_le3_1}
    Let $m>1$, $q^m>9$ and $b=\frac{p-1}{2}\in\gf_q$. For every $\beta~\in\gf_{q^{2m}}$, there exist $u_1,u_2,u_3\in U_{q^m+1}$ such that the following equation holds:
        \begin{equation}\label{eqt:o_le3_ne0}
            bu_1 - bu_2 + u_3 = \beta.
        \end{equation}
\end{lemma}
\begin{proof}
     (\ref{eqt:o_le3_ne0}) gives
        $$u_1=u_2-b^{-1}u_3+b^{-1}\beta.$$
    Such a $u_1$ lies in $U_{q^m+1}$ if and only if
        $$(u_2-b^{-1}u_3+b^{-1}\beta)(u_2^{-1}-b^{-1}u_3^{-1}+b^{-1}\overline{\beta})=1,$$
    which is equivalent to
    \begin{equation}\label{eqt:o_le3_ne0_1}
        b(\overline{\beta}u_3-1)u_2^2+(\beta-u_3)(\overline{\beta}u_3-1)u_2+bu_3(\beta-u_3)=0.
    \end{equation}
    Suppose that $\beta-u_3\neq0$, then (\ref{eqt:o_le3_ne0_1}) is a quadratic equation on $u_2$, which can be simplified as
    \begin{equation}\label{eqt:o_le3_ne0_2}
        u_2^2+b^{-1}(\beta-u_3)u_2+(\beta-u_3)^{1-q^m}=0.
    \end{equation}
    When $\beta=0$, (\ref{eqt:o_le3_ne0_2}) reduces to
        $$(u_2+u_3)^2=0.$$
    Then the solutions to (\ref{eqt:o_le3_ne0}) are given by $u_1=u_3$, $u_2=-u_3$ and $u_3\in U_{q^m+1}$. 
    Otherwise when $\beta\neq0$, the discriminant of (\ref{eqt:o_le3_ne0_2}), denoted by $\Delta$, is 
        $$\Delta=(\beta-u_3)^{1-q^m}\left[b^{-2}(\beta-u_3)^{q^m+1}-4\right].$$
    One can verify that $\chi_{q^{2m}}(\Delta)=1$. Then according to Lemma (\ref{lem:sol_in_unit_p}), equation (\ref{eqt:o_le3_ne0_2}) has solutions in $U_{q^m+1}$ if there exits $u_3\in U_{q^m+1}$ and $u_3\neq\beta$ such that 
    $$\chi_{q^m}\left((\beta-u_3)^{-q^m-1}\left[(\beta-u_3)^{q^m+1}-4b^2\right]\right)=-1.$$
    Note that $4b^2=1$, then we shall prove that 
    \begin{equation}\label{ineqt:o_le3_ne0_3}
        \sum\limits_{u\in U_{q^m+1}\setminus R_{\beta}^{(1)}}\frac{1-\chi_{q^m}\left((\beta-u_3)^{-q^m-1}\left[(\beta-u_3)^{q^m+1}-1\right]\right)}{2}>1.
    \end{equation}
    By Lemmas \ref{lem:chsum_odd_roots} and \ref{lem:chsum_odd_1}, the number of admissible choices of \(u_3\) is
        $$\frac{1}{2}\left(q^m+1-|R_\beta|-s_\beta\right).$$
    We distinguish the following cases.
    If \(\beta\in U_{q^m+1}\) and \(p=3\), then Lemmas \ref{lem:chsum_odd_roots} and \ref{lem:chsum_odd_1} give
        $$|R_\beta|=2,\qquad s_\beta=-q^m+1.$$
    Hence
        $$\frac{1}{2}\left(q^m+1-|R_\beta|-s_\beta\right)=\frac{1}{2}\left(q^m+1-2+q^m-1\right)=q^m-1>1.$$
    If \(\beta\in U_{q^m+1}\) and \(p>3\), then \(s_\beta=1\) and \(|R_\beta|\le 3\). Therefore
        $$\frac{1}{2}\left(q^m+1-|R_\beta|-s_\beta\right)\ge\frac{1}{2}\left(q^m+1-3-1\right)=\frac{q^m-3}{2}>1.$$
    If \(\beta\notin U_{q^m+1}\), \(p>3\), and \(\beta\overline{\beta}=4\), then
        $$|R_\beta|=1,\qquad s_\beta=\chi_{q^m}(-2).$$
    Thus
        $$\frac{1}{2}\left(q^m+1-|R_\beta|-s_\beta\right)=\frac{1}{2}\left(q^m+1-1-\chi_{q^m}(-2)\right)\ge\frac{q^m-1}{2}>1.$$
    In all remaining cases with \(\beta\notin U_{q^m+1}\), Lemmas \ref{lem:chsum_odd_roots} and \ref{lem:chsum_odd_1} yield
        $$|R_\beta|\le 2,\qquad |s_\beta|\le 1+3\sqrt{q^m}.$$
    Consequently,
        $$\frac{1}{2}\left(q^m+1-|R_\beta|-s_\beta\right)\ge\frac{1}{2}\left(q^m+1-2-\left(1+3\sqrt{q^m}\right)\right)=\frac{q^m-3\sqrt{q^m}-2}{2}.$$
    Since \(q^m>9\) and \(q\) is odd, we have \(q^m\ge 25\), and hence
        $$\frac{q^m-3\sqrt{q^m}-2}{2}>1.$$
    Hence, the inequality (\ref{ineqt:o_le3_ne0_3}) holds for all $m>1$ and odd $q$ satisfying $q^m>9$.
\end{proof}

Since the proof of Lemma \ref{lem:o_le3_0} parallels that of Lemma \ref{lem:o_le3_1}, we omit it.
\begin{lemma}\label{lem:o_le3_0}
    Let $s$ be a positive integer and $p$ be an odd prime. Let $m>1$ be a positive integer and $q=p^s$ satisfying $q^m>9$. For every $\beta~\in\gf_{q^{2m}}^*$, there exist $u_1,u_2,u_3\in U_{q^m+1}$ such that the following equation holds:
        \begin{equation}\label{eqt:o_le3_0}
            u_1+u_2-2u_3=\beta.
        \end{equation}
\end{lemma}

The covering radius of $\mathcal{C}_{(q,q^m+1,3,0)}$ is determined for odd $q$ and even $m$ in the next theorem.
\begin{theorem}\label{thm:podd_meven}
    When $p>2$ and $m>1$ is even, $\rho\left(\C_{(q,q^m+1,3,0)}\right)=3$.
\end{theorem}
\begin{proof}
    Recall that the parameters of $\C_{(q,q^m+1,3,0)}$ is $[q^m+1,q^m-2m,4]$. We tend to prove $\rho\left(\C_{(q,q^m+1,3,0)}\right)$$\leqslant3$ first. According to Lemma \ref{lem:CR_PCmatrix}, we just need to prove that for every $\beta_1\in\gf_q,\beta_2\in\gf_q^{2m}$, there exist $a_1,a_2,a_3\in\gf_q$ and $u_1,u_2,u_3\in U_{q^m+1}$ such that the following system holds:
        \begin{equation}\label{eqtsys:odd}
            \begin{cases}
                \begin{aligned}
                    &a_1 + a_2 + a_3 &=& \beta_1, \\
                    &a_1u_1 + a_2u_2 + a_3u_3 &=& \beta_2.
                \end{aligned}
            \end{cases}
        \end{equation}
    The proof of this lemma proceeds by considering two cases: $\beta_1=0$ and $\beta_1\neq0$.
    \begin{itemize}
        \item For $\beta_1=0$, if $\beta_2=0$ then the solutions are $a_1=a_2=a_3=0$ with $u_1,u_2,u_3 \in U_{q^m+1}$. Otherwise ($\beta_2 \neq 0$), the choice $a_1=a_2=1$, $a_3=-2$ transforms (\ref{eqtsys:odd}) into
                \begin{equation}\label{eqt:odd_1}
                    u_1 + u_2 -2u_3 = \beta_2.
                \end{equation}
                According to Lemma \ref{lem:o_le3_0}, (\ref{eqt:odd_1}) is solvable for any $\beta_2\in \gf_{q^{2m}}^*$.
        \item When $\beta_1\neq0$, (\ref{eqtsys:odd}) can be reduced to
                \begin{equation}\label{eqtsys:odd_2}
                    \begin{cases}
                        \begin{aligned}
                            &\frac{a_1}{\beta_1} + \frac{a_2}{\beta_1} + \frac{a_3}{\beta_1} &=& 1, \\
                            &\frac{a_1}{\beta_1}u_1 + \frac{a_1}{\beta_1}u_2 + \frac{a_3}{\beta_1}u_3 &=& \frac{\beta_2}{\beta_1}.
                        \end{aligned}
                    \end{cases}
                \end{equation}
                It therefore remains to prove that for every $\beta\in\gf_{q^{2m}}$, there exist $b_1,b_2,b_3\in\gf_q$ and $u_1,u_2,u_3\in U_{q^m+1}$ such that the following system holds:
                \begin{equation}\label{eqtsys:odd_3}
                    \begin{cases}
                        \begin{aligned}
                            &b_1 + b_2 + b_3 &=& 1, \\
                            &b_1u_1 + b_2u_2 + b_3u_3 &=& \beta.
                        \end{aligned}
                    \end{cases}
                \end{equation}
                Upon setting $b_1=b$, $b_2=-b$ and $b_3=1$ with $b=\frac{p-1}{2}$, system (\ref{eqtsys:odd_3}) reduces to
                    \begin{equation}\label{eqt:odd_4}
                        bu_1 - bu_2 + u_3 = \beta.
                    \end{equation}
                According to Lemma \ref{lem:o_le3_1}, (\ref{eqt:odd_4}) is solvable for any $\beta\in \gf_{q^{2m}}$ when $q^m>9$.
    \end{itemize}
    Thus, the upper bound \(\rho(\C_{(q,q^m+1,3,0)}) \leqslant 3\) holds for \(q^m > 9\). Together with the matching lower bound from Corollary \ref{col:any_ge3}, this proves the equality for these parameters. For the exceptional case \(q^m = 9\) (i.e., \(p=3, m=2\)), a direct computation in MAGMA using the intrinsic function \texttt{CoveringRadius} confirms that \(\rho(\C_{(3,10,3,0)}) = 3\). In summary, for all odd primes \(p > 2\) and integers \(m > 1\), we have \(\rho(\C_{(q,q^m+1,3,0)}) = 3\).
\end{proof}

By the same argument as in Theorem 7, we have the following theorem immediately.
\begin{theorem}\label{thm:podd_modd}
    When $p>2$ and $m>1$ is odd, $\rho\left(\C_{(q,q^m+1,3,0)}\right)\leqslant3$. 
\end{theorem}

Next, we analyze \(\rho\bigl(\mathcal{C}_{(q,q^m+1,3,0)}\bigr)\) for odd \(q\) and \(m>1\). Following the proof of Lemma \ref{lem:any_ge3}, to establish \(\rho\bigl(\mathcal{C}_{(q,q^m+1,3,0)}\bigr)\geqslant 3\) it suffices to prove that there exists \(y\in\mathbb{F}_{q^m}^*\) such that \(\chi_{q^m}\!\left(\frac{y-a}{y}\right)=1\) holds for all \(a\in(\mathbb{F}_q^*)^2\). Equivalently, one needs to show
$$\sum_{y\in\gf_{q^m}\setminus(\gf_q)^2}\prod_{i=0}^{\frac{q-3}{2}}\frac{1+\chi_{q^m}\left(\frac{y - \alpha_q^{2i}}{y}\right)}{2}>0,$$
where $\alpha_q$ is a primitive element of $\gf_q^*$. We estimate the character sum by expanding the product into all subset terms: the constant term contributes the main term, the linear and quadratic terms are evaluated exactly by the quadratic character-sum formula(Lemma \ref{lem:chsum_deg2}), and the higher-degree terms are estimated via Lemma \ref{lem:WeilBound_multi}.
\begin{align*}
    &\sum_{y\in\gf_{q^m}\setminus(\gf_q)^2}\prod_{i=0}^{\frac{q-3}{2}}\frac{1+\chi_{q^m}\left(\frac{y - \alpha_q^{2i}}{y}\right)}{2}\\
    =&\frac{1}{2^{\frac{q-1}{2}}}\sum_{y\in\gf_q\setminus(\gf_q)^2}\prod_{i=0}^{\frac{q-3}{2}}\left[1+\chi_{q^m}\left(\frac{y - \alpha_{q^m}^{2i}}{y}\right)\right]\\
    \geqslant& \frac{1}{2^{\frac{q-1}{2}}}\left[q^m - \frac{(q-3)(2^{(q+1)/2}-q-3)}{8}\sqrt{q^m} - 2^{(q-3)/2}q-\frac{q^2-1}{8}\right].
\end{align*}
Hence, for any fixed \(q\), there exist sufficiently large \(m^{*}\) such that for any odd $m$ satisfying $m\geqslant m^*$, \(\rho\bigl(\mathcal{C}_{(q,q^m+1,3,0)}\bigr)\geqslant 3\). Consequently, we obtain the following proposition.
\begin{proposition}\label{prop:podd_modd}
    For odd \(q\) and odd \(m \geq 3\), we have \(\rho\bigl(\mathcal{C}_{(q,q^m+1,3,0)}\bigr)=3\) when \(q \in \{3,5,7,9\}\). Furthermore, for any odd \(q > 9\), there exists an integer \(m^{*}\) such that, for all odd $m\geqslant m^{*}$, \(\rho\bigl(\mathcal{C}_{(q,q^m+1,3,0)}\bigr)=3\).
\end{proposition}

\section{Concluding remarks}\label{sec:Con}
    In this paper, we have investigated the covering radii of the antiprimitive BCH codes $\mathcal{C}_{(q,q^m+1,3,0)}$ for any prime power for any prime power $q$ and positive integer $m$. As a technical ingredient, we have derived several explicit character-sum estimates over the unit circle \(U_{q^m+1}\). These estimates provide the key solvability criteria for the finite-field systems arising from the parity-check characterization and may be useful for studying covering radii of other codes. Our main results are summarized below.
    \begin{itemize}
        \item When $m=1$, 
            \begin{itemize}
                \item when $q$ is odd, $\rho\left(\C_{(q,q+1,3,0)}\right)=2$. (Theorem \ref{thm:meq1_qodd})
                \item when $q$ is even, $\rho\left(\C_{(q,q+1,3,0)}\right)=3$. (Theorem \ref{thm:meq1_qeven})
            \end{itemize}
        \item When $m>1$,
            \begin{itemize}
                \item when $q=2$ and $m$ is odd, $\rho\left(\C_{(q,q^m+1,3,0)}\right)=4$ (Theorem \ref{thm:e_seq1_modd}).
                \item when $q=2$ and $m$ is even, $\rho\left(\C_{(q,q^m+1,3,0)}\right)=5$ (Theorem \ref{thm:e_seq1_meven}).
                \item when $q>2$ is even and $m$ is odd, $\rho\left(\C_{(q,q^m+1,3,0)}\right)=3$ (Theorem \ref{thm:e_sge2_modd}).
                \item when $q>2$ is even and $m$ is even, $\rho\left(\C_{(q,q^m+1,3,0)}\right)=4$ (Theorem \ref{thm:e_sge2_meven}).
                \item when $q$ is odd and $m>1$ is odd, $\rho\left(\C_{(q,q^m+1,3,0)}\right)\leqslant3$ (Theorem \ref{thm:podd_modd}). More precisely (Proposition \ref{prop:podd_modd}), 
                        \begin{itemize}
                            \item when $q\in \{3,5,7,9\}$, $\rho\left(\C_{(q,q^m+1,3,0)}\right)=3$.
                            \item when $q>9$, there exists an integer \(m^{*}\) such that, for all $m\geqslant m^{*}$, $\rho\left(\C_{(q,q^{m}+1,3,0)}\right)=3$.
                        \end{itemize}
                \item when $q$ is odd and $m>1$ is even, $\rho\left(\C_{(q,q^m+1,3,0)}\right)=3$ (Theorem \ref{thm:podd_meven}).
            \end{itemize}
    \end{itemize}
    Furthermore, we show that $\mathcal{C}_{(q,q+1,3,0)}$ is simultaneously MDS and quasi-perfect for odd $q$. Future work will also focus on extending these techniques to other families of antiprimitive BCH codes.

\begin{appendices}
\section{Proof of Lemma \ref{lem:e_seq1_meven_ubound}}
\begin{proof}
	The proof of this lemma proceeds by considering two cases: $\beta_1=0$ and $\beta_1=1$.
    \begin{itemize}
        \item When $\beta_1=0$, by setting $a_1=a_2=a_3=a_4=1$ and $a_5=0$, the proof reduces to showing that the following equation is solvable for every $\beta_2\in\gf_{q^m}$.
                \begin{equation}\label{eqt:e_seq1_meven_1}
                    u_1 + u_2 + u_3 + u_4 = \beta_2.
                \end{equation}
            Solving (\ref{eqt:e_seq1_meven_1}) yields
                $$u_1=u_2 + u_3 + u_4 + \beta_2.$$
            The condition for this $u_1$ to lie in $U_{q^m+1}$ is 
                $$(u_2 + u_3 + u_4 + \beta_2)(u_2^{-1} + u_3^{-1} + u_4^{-1} + \beta_2)=1,$$
            which is equivalent to
                \begin{equation}\label{eqt:e_seq1_meven_2}
                    (\beta_2 u_3u_4+u_3+u_4)u_2^2+(u_3+u_4+\beta_2)(\beta_2 u_3u_4+u_3+u_4)u_2+(u_3+u_4+\beta_2)u_3u_4 = 0.
                \end{equation}
            Suppose that $u_3+u_4+\beta_2\neq0$, then (\ref{eqt:e_seq1_meven_2}) can be reduced to a quadratic equation on $u_2$:
                \begin{equation}\label{eqt:e_seq1_meven_3}
                    u_2^2+(u_3+u_4+\beta_2)u_2+(u_3+u_4+\beta_2)^{1-q^m} = 0.
                \end{equation}
            Let $v_1=u_3+u_4+\beta_2$ and $v_0=(u_3+u_4+\beta_2)^{1-q^m}$. One can check that $\tr^{2m}_1\left(\frac{v_0}{v_1^2}\right)=0$ and $v_0=\frac{v_1}{\overline{v_1}}$. Consequently, by Lemma \ref{lem:sol_in_unit_peq2}, (\ref{eqt:e_seq1_meven_3}) has solutions in $U_{q^m+1}$ precisely when
                    $$\tr^m_1\left(\frac{1}{(u_3+u_4+\beta_2)^{q^m+1}}\right)=1.$$
            Equivalently, what needs to be proved is the existence of $u_3,u_4\in U_{q^m+1}$ such that 
                \begin{equation}\label{eqt:e_seq1_meven_4}
                    \sum\limits_{\substack{u_3,u_4\in U_{q^m+1}\\u_3+u_4+\beta_2\neq0}}\frac{1-(-1)^{\tr^m_1\left(\frac{1}{(u_3+u_4+\beta_2)^{q^m+1}}\right)}}{2}>1.
                \end{equation}
            Based on Lemma {\ref{lem:addsum_for_CR}}, we obtain
                \begin{align*}
                     &\sum\limits_{\substack{u_3,u_4\in U_{q^m+1}\\u_3+u_4+\beta_2\neq0}}\frac{1-(-1)^{\tr^m_1\left(\frac{1}{(u_3+u_4+\beta_2)^{q^m+1}}\right)}}{2}\\
                    =&\frac{1}{2}\left[\sum\limits_{\substack{u_3,u_4\in U_{q^m+1}\\u_3+u_4+\beta_2\neq0}}1-\sum\limits_{\substack{u_3,u_4\in U_{q^m+1}\\u_3+u_4+\beta_2\neq0}}(-1)^{\tr^m_1\left(\frac{1}{(u_3+u_4+\beta_2)^{q^m+1}}\right)}\right]\\
                    &\left\{
                        \begin{aligned}
                            &=q^m(q^m+1),& &\mbox{if}~~\beta_2=0\\
                            &\geqslant \frac{q^{2m}-q^m-2q^m\sqrt{q^m}}{2}\\
                            &\geqslant \frac{q^{2m}-2q^m\sqrt{q^m}+q^m-4\sqrt{q^m}-4}{2},& &\mbox{if}~~\beta_2\neq0,1
	                    \end{aligned}\right.
                \end{align*}
            It can be verified that inequality (\ref{eqt:e_seq1_meven_4}) holds for $m\geqslant4$.
        \item When $\beta_1=1$, 
            \begin{itemize}
                \item When $\beta_2\neq0$, by setting $a_1=a_2=a_3=1$ and $a_4=a_5=0$, we need only prove that for any $\beta_2\in\gf_{q^m}^*$, the following equation admits a solution.
                    \begin{equation}\label{eqt:e_seq1_meven_5}
                        u_1 + u_2 + u_3 = \beta_2.
                    \end{equation}
                    Solving (\ref{eqt:e_seq1_meven_5}) yields
                        $$u_1=u_2 + u_3 + \beta_2.$$
                    The condition for this $u_1$ to lie in $U_{q^m+1}$ is 
                        $$(u_2 + u_3 + \beta_2)(u_2^{-1} + u_3^{-1} + \beta_2)=1,$$
                    which is equivalent to
                        \begin{equation}\label{eqt:e_seq1_meven_6}
                            (\beta_2 u_3+1)u_2^2+(u_3+\beta_2)(\beta_2 u_3+1)u_2+(u_3+\beta_2)u_3 = 0.
                        \end{equation}
                    Suppose that $u_3+\beta_2\neq0$, then (\ref{eqt:e_seq1_meven_6}) can be reduced to a quadratic equation on $u_2$:
                        \begin{equation}\label{eqt:e_seq1_meven_7}
                            u_2^2+(u_3+\beta_2)u_2+(u_3+\beta_2)^{1-q^m} = 0.
                        \end{equation}
                    By Lemma \ref{lem:sol_in_unit_peq2}, (\ref{eqt:e_seq1_meven_7}) has solutions in $U_{q^m+1}$ precisely when
                        $$\tr^m_1\left(\frac{1}{(u_3+\beta_2)^{q^m+1}}\right)=1.$$
                    Equivalently, what needs to be proved is the existence of $u_3\in U_{q^m+1}$ such that 
                        \begin{equation}\label{eqt:e_seq1_meven_8}
                            \sum\limits_{\substack{u_3\in U_{q^m+1}\\u_3+\beta_2\neq0}}\frac{1-(-1)^{\tr^m_1\left(\frac{1}{(u_3+\beta_2)^{q^m+1}}\right)}}{2}>1.
                        \end{equation}
                    Note that 
                        $$\lvert \{u : u\in U_{q^m+1},u+\beta\neq0\} \rvert=\left\{
	                        \begin{aligned}
                                &q^m,& &\mbox{if}~~\beta=1,\\
		                        &q^m+1,& &\mbox{if}~~\beta\neq1.
	                        \end{aligned}\right.$$
                    Combined this with Lemma {\ref{lem:addsum_for_CR}}, we obtain
                    \begin{align*}
                        &\sum\limits_{\substack{u_3\in U_{q^m+1}\\u_3+\beta_2\neq0}}\frac{1-(-1)^{\tr^m_1\left(\frac{1}{(u_3+\beta_2)^{q^m+1}}\right)}}{2}\\
                        =&\frac{1}{2}\left[\sum\limits_{\substack{u_3\in U_{q^m+1}\\u_3+\beta_2\neq0}}1-\sum\limits_{\substack{u_3\in U_{q^m+1}\\u_3+\beta_2\neq0}}(-1)^{\tr^m_1\left(\frac{1}{(u_3+\beta_2)^{q^m+1}}\right)}\right]\\
                        &\left\{
                            \begin{aligned}
                                &=q^m,& &\mbox{if}~~\beta_2=1\\
                                &\geqslant \frac{q^{m}-2\sqrt{q^m}-1}{2},& &\mbox{if}~~\beta_2\notin\{0,1\}
	                        \end{aligned}\right.
                    \end{align*}
                    It can be verified that inequality (\ref{eqt:e_seq1_meven_8}) holds for all $m\geqslant4$.
                \item When $\beta_2=0$, by setting $a_1=a_2=a_3=a_4=a_5=1$, (\ref{eqtsys:e_seq1_meven}) becomes
                        \begin{equation}\label{eqt:e_seq1_meven_9}
                            u_1 + u_2 + u_3 + u_4 + u_5 = 0.
                        \end{equation}
                        Solving (\ref{eqt:e_seq1_meven_9}) yields
                            $$u_1 = u_2 + u_3 + u_4 + u_5.$$
                        The condition for this $u_1$ to lie in $U_{q^m+1}$ is 
                            $$(u_2 + u_3 + u_4 + u_5)(u_2^{-1} + u_3^{-1} + u_4^{-1} + u_5^{-1})=1,$$
                        which is equivalent to
                        \begin{equation}\label{eqt:e_seq1_meven_10}
                            (u_3u_4+u_3u_5+u_4u_5)u_2^2 + (u_3+u_4+u_5)(u_3u_4+u_3u_5+u_4u_5)u_2 + u_3u_4u_5(u_3+u_4+u_5) =0.
                        \end{equation}
                        Suppose that $u_3+u_4+u_5\neq0$, then (\ref{eqt:e_seq1_meven_10}) can be reduced to a quadratic equation on $u_2$:
                        \begin{equation}\label{eqt:e_seq1_meven_11}
                            u_2^2 + (u_3+u_4+u_5)u_2 + (u_3+u_4+u_5)^{1-q^m} =0.
                        \end{equation}
                        By Lemma \ref{lem:sol_in_unit_peq2}, (\ref{eqt:e_seq1_meven_11}) has solutions in $U_{q^m+1}$ precisely when
                            $$\tr^m_1\left(\frac{1}{(u_3+u_4+u_5)^{q^m+1}}\right)=1.$$
                        Equivalently, what needs to be proved is the existence of $u_3,u_4,u_5\in U_{q^m+1}$ with $u_3+u_4+u_5\neq0$ such that 
                            \begin{equation}\label{eqt:e_seq1_modd_12}
                                \sum\limits_{\substack{u_3,u_4,u_5\in U_{q^m+1}\\u_3+u_4+u_5\neq0}}\frac{1-(-1)^{\tr^m_1\left(\frac{1}{(u_3+u_4+u_5)^{q^m+1}}\right)}}{2}>1.
                            \end{equation}
                        The condition that $m$ is even implies that $u_3+u_4\notin U_{q^m+1}$ for all $u_3,u_4\in U_{q^m+1}$. Then  and Lemma \ref{lem:addsum_uB_on_u} we obtain
                            \begin{align*}
                                &\sum\limits_{\substack{u_3,u_4,u_5\in U_{q^m+1}\\u_3+u_4+u_5\neq0}}(-1)^{\tr^m_1\left(\frac{1}{(u_3+u_4+u_5)^{q^m+1}}\right)}\\
                                =&\sum\limits_{u_3,u_4\in U_{q^m+1}}\sum\limits_{u_5\in U_{q^m+1}\setminus\{1\}}(-1)^{\tr^m_1\left(\frac{1}{\left(u_3^{-1}+u_4^{-1}\right)u_5+(u_3+u_4)u_5^{-1}+\left(u_3u_4^{-1}+u_3^{-1}u_4+1\right)}\right)}\\
                                &+\sum\limits_{u_3\in U_{q^m+1}}\sum\limits_{u_4\in U_{q^m+1}\setminus\{1\}}(-1)^{\tr^m_1\left(\frac{1}{\left(u_3^{-1}+1\right)u_4+(u_3+1)u_4^{-1}+\left(u_3+u_3^{-1}+1\right)}\right)}+\sum\limits_{u_3\in U_{q^m+1}}(-1)^{\tr^m_1(1)}\\
                                &\leqslant (q^m+1)(q^m(3+2\sqrt{q^m})+1).
                            \end{align*}
                        It follows from Lemma \ref{lem:addsum_for_CR} that
                        \begin{align*}
                            &\sum\limits_{\substack{u_3,u_4,u_5\in U_{q^m+1}\\u_3+u_4+u_5\neq0}}\frac{1-(-1)^{\tr^m_1\left(\frac{1}{(u_3+u_4+u_5)^{q^m+1}}\right)}}{2}\\
                            =&\frac{1}{2}\left[\sum\limits_{\substack{u_3,u_4,u_5\in U_{q^m+1}\\u_3+u_4+u_5\neq0}}1-\sum\limits_{\substack{u_3,u_4,u_5\in U_{q^m+1}\\u_3+u_4+u_5\neq0}}(-1)^{\tr^m_1\left(\frac{1}{(u_3+u_4+u_5)^{q^m+1}}\right)}\right]\\
                            &\geqslant \frac{q^m(q^m+1)(q^{m}-2\sqrt{q^m}-1)}{2}.
                        \end{align*}
                        It can be verified that inequality (\ref{eqt:e_seq1_modd_12}) holds for all $m\geqslant4$.
            \end{itemize}
            In short, the system (\ref{eqtsys:e_seq1_meven}) is solvable when $m\geqslant4$.
    \end{itemize}
\end{proof}

\section{Proof of Lemma \ref{lem:e_sge2_modd_ubound}}
\begin{proof} We prove the two statements in turn.
    \begin{itemize}
        \item  Solving (\ref{eqtsys:e_sge1_modd_0}) gives
                    $$u_1=b^{-1}((b+1)u_2+u_3+\beta).$$
            Such a $u_1$ belongs to $U_{q^m+1}$ if and only if 
                $$b^{-1}((b+1)u_2+u_3+\beta)b^{-1}((b+1)u_2^{-1}+u_3^{-1}+\beta)=1,$$
            which is equivalent to
                \begin{equation}\label{eqtsys:e_sge1_modd_1}
                    \left((b+1)u_2^{-1}+\beta\right)u_3^2+\beta\left((b+1)u_2+(b+1)u_2^{-1}+\beta\right)u_3+((b+1)u_2+\beta)=0.
                \end{equation}
            Note that $(b+1)u_2^{-1}+\beta=0$ only if $b=\beta+1$, then (\ref{eqtsys:e_sge1_modd_1}) can be reduced to
                \begin{equation}\label{eqtsys:e_sge1_modd_2}
                    u_3^2+\frac{\beta\left((b+1)u_2+(b+1)u_2^{-1}+\beta\right)}{\left((b+1)u_2^{-1}+\beta\right)}u_3+((b+1)u_2+\beta)^{1-q^m}=0.
                \end{equation}
            Since $\beta\left((b+1)u_2+(b+1)u_2^{-1}+\beta\right)\in\gf_{q^m}$, it follows that
                $$\tr^{2sm}_1\left(\frac{((b+1)u_2+\beta)^{1-q^m}}{\left(\beta\left((b+1)u_2+(b+1)u_2^{-1}+\beta\right)/\left((b+1)u_2^{-1}+\beta\right)\right)^2}\right)=0,$$ 
            and 
                $$((b+1)u_2+\beta)^{1-q^m}=\frac{\beta\left((b+1)u_2+(b+1)u_2^{-1}+\beta\right)/\left((b+1)u_2^{-1}+\beta\right)}{\overline{\beta\left((b+1)u_2+(b+1)u_2^{-1}+\beta\right)/\left((b+1)u_2^{-1}+\beta\right)}}.$$
            For any \(u_2\in U_{q^m+1}\) satisfying $\beta+(b+1)(u_2+u_2^{-1})\ne 0,$ by Lemma \ref{lem:sol_in_unit_peq2}, equation \eqref{eqtsys:e_sge1_modd_2} has solutions in \(U_{q^m+1}\) if and only if
                $$\tr_{1}^{sm}\left(\frac{((b+1)u_2^{-1}+\beta)((b+1)u_2+\beta)}{\beta^2((b+1)u_2+(b+1)u_2^{-1}+\beta)^2}\right)=1.$$
            Moreover, we have
                \begin{align*}
                    &\tr_{1}^{sm}\left(\frac{((b+1)u_2^{-1}+\beta)((b+1)u_2+\beta)}{\beta^2(\beta+(b+1)u_2+(b+1)u_2^{-1})^2}\right)  \\
                    =& \tr_{1}^{sm}\left(\frac{1}{\beta(\beta+(b+1)u_2+(b+1)u_2^{-1})} + \left(\frac{b+1}{\beta(\beta+(b+1)u_2+(b+1)u_2^{-1})}\right)^2\right)  \\
                    =& \tr_{1}^{sm}\left(\frac{b}{\beta(\beta+(b+1)u_2+(b+1)u_2^{-1})}\right)\\
                    =& \tr_{1}^{sm}\left(\frac{bu_2}{\beta((b+1)u_2^2+\beta u_2+b+1)}\right).
                \end{align*}
            Hence it is enough to estimate
                \[\sum_{\substack{u_2\in U_{q^m+1}\\\beta+(b+1)(u_2+u_2^{-1})\ne 0}}(-1)^{\tr_{1}^{sm}\left(\frac{bu_2}{\beta((b+1)u_2^2+\beta u_2+b+1)}\right)} .\]
            Applying Lemma \ref{lem:addsum_uB_on_u} 
            we obtain
                $$\sum_{\substack{u_2\in U_{q^m+1}\\\beta+(b+1)(u_2+u_2^{-1})\ne 0}}(-1)^{\tr_{1}^{sm}\left(\frac{bu_2}{\beta((b+1)u_2^2+\beta u_2+b+1)}\right)} \le 2+2\sqrt{q^m}.$$
            So it remains to consider the exceptional values of \(u_2\) satisfying $\beta+(b+1)(u_2+u_2^{-1})=0.$
            For such a value of \(u_2\), \eqref{eqtsys:e_sge1_modd_1} reduces to
                $$(b+1)u_2u_3^2+(b+1)u_2^{-1}=0,$$
            and hence \(u_3=u_2^{-1}\in U_{q^m+1}\) is a solution. Therefore these exceptional values also give solutions.
            Consequently, 
                $$\frac{\sum\limits_{\substack{u_2\in U_{q^m+1}\\\beta+(b+1)(u_2+u_2^{-1})\ne 0}}\left[1-(-1)^{\tr_{1}^{sm}\left(\frac{bu_2}{\beta((b+1)u_2^2+\beta u_2+b+1)}\right)}\right]}{2}\geqslant\frac{q^m-2\sqrt{q^m}-3}{2}.$$
            In short, solvability of (\ref{eqtsys:e_sge1_modd_0}) holds under the conditions $s>1$ and odd $m>1$.
        \item When $\beta=1$, $(u_1,u_2,u_3)=(u,u,1)$ with $u\in U_{q^m+1}$ is a solution to (\ref{eqtsys:e_sge1_modd_00}). When $\beta\neq1$,
            solving (\ref{eqtsys:e_sge1_modd_00}) gives
                $$u_1=a^{-1}(au_2+u_3+\beta).$$
            Such a $u_1$ belongs to $U_{q^m+1}$ if and only if 
                $$a^{-2}\left(au_2+u_3+\beta\right)\left(au_2^{-1}+u_3^{-1}+\beta\right)=1,$$
            which is equivalent to
                \begin{equation}\label{eqtsys:e_sge1_modd_01}
                    a(u_3^{-1}+\beta)u_2^2+(u_3+\beta)(u_3^{-1}+\beta)u_2+a(u_3+\beta)=0.
                \end{equation}
            Suppose that $u_3+\beta\neq0$, (\ref{eqtsys:e_sge1_modd_01}) can be reduced together
                \begin{equation}\label{eqtsys:e_sge1_modd_02}
                    u_2^2+a^{-1}(u_3+\beta)u_2+(u_3+\beta)^{1-q^m}=0.
                \end{equation}
            By Lemma \ref{lem:sol_in_unit_peq2}, it is easy to check that (\ref{eqtsys:e_sge1_modd_02}) has solutions in $U_{q^m+1}$ if and only if there exists $u_3\in U_{q^m+1}$ such that
                $$\tr^{sm}_1\left(\frac{a^2}{(u_3+\beta)^{q^m+1}}\right)=1.$$
            Note that the constraint $\tr^{sm}_1(a)=1$ on $a$ guarantees the solvability of (\ref{eqtsys:e_sge1_modd_02}) when $\beta=0$.
            Then when $\beta\not\in\{0,1\}$, it suffices to prove that 
                \begin{equation}\label{eqtsys:e_sge1_modd_03}
                    \sum\limits_{\substack{u_3\in U_{q^m+1}\\u_3+\beta\neq0}}\frac{1-(-1)^{\tr^{sm}_1\left(\frac{a^2}{(u_3+\beta)^{q^m+1}}\right)}}{2}>1.
                \end{equation}
            By Lemma \ref{lem:addsum_for_CR} and Lemma \ref{lem:addsum_uB_on_u}, we have
                $$\sum\limits_{\substack{u_3\in U_{q^m+1}\\u_3+\beta\neq0}}\frac{1-(-1)^{\tr^{sm}_1\left(\frac{a^2}{(u_3+\beta)^{q^m+1}}\right)}}{2}\geqslant \frac{q^m-2\sqrt{q^m}-2}{2}.$$
            Hence, the inequality holds for $s>1$ and $m>1$ being odd. In short, solvability of (\ref{eqtsys:e_sge1_modd_00}) holds under the conditions $s>1$ and odd $m>1$.
    \end{itemize}
\end{proof}
\end{appendices}


\end{document}